\documentclass[12pt]{article}

\newcommand{\agwb}{A_{\mathrm{gwb}}}

\newcommand{\comment}[1]{}

\usepackage{scicite}
\usepackage{graphicx}
\usepackage{times}
\usepackage{textcomp}
\usepackage{amssymb}
\usepackage[left]{lineno}

\usepackage[multiple]{footmisc}

\newenvironment{sciabstract}{%
\begin{quote} \bf}
{\end{quote}}

\title{A Gamma-ray Pulsar Timing Array Constrains the Nanohertz Gravitational Wave Background}

\author{The Fermi-LAT Collaboration$^{*,\dagger}$}

\date{}

\begin{document} 

\baselineskip24pt

\maketitle 
\noindent



\begin{sciabstract}

\vspace{-0.5cm}

After large galaxies merge, their central supermassive black holes are expected to form binary systems whose orbital motion generates a gravitational wave background (GWB) at nanohertz frequencies. Searches for this background utilize pulsar timing arrays, which perform long-term monitoring of millisecond pulsars (MSPs) at radio wavelengths.  We use 12.5 years of Fermi Large Area Telescope data to form a gamma-ray pulsar timing array.  Results from 35 bright gamma-ray pulsars place a 95\% credible limit on the GWB characteristic strain of $1.0\times10^{-14}$ at 1 yr$^{-1}$, which scales as the observing time span $t_{\mathrm{obs}}^{-13/6}$. This direct measurement provides an independent probe of the GWB while offering a check on radio noise models.

\end{sciabstract}
$*$ {\footnotesize Fermi-LAT Collaboration authors and affiliations are listed in the supplementary materials.}\\
$\dagger$ {\footnotesize Corresponding Authors: Matthew Kerr (matthew.kerr@gmail.com), Aditya Parthasarathy (adityapartha3112@gmail.com)}
\pagebreak


\section*{Main Text}

Pulsars are spinning neutron stars that emit beams of broadband radiation from radio to gamma-ray wavelengths that appear to pulse as they periodically sweep across the line of sight to Earth \cite{Hewish68}.  Millisecond pulsars (MSPs) spin at hundreds of hertz and pulse with sufficient regularity to function as celestial clocks distributed across the sky and throughout the Galaxy.
Timing of individual MSPs using radio telescopes has been used to test General Relativity and alternate theories of gravity \cite{Will14}. Long-term monitoring campaigns of ensembles of MSPs are used to search for low-frequency gravitational waves, expected from supermassive black hole (SMBH) binaries that are predicted to exist at the centers of galaxies that have undergone mergers. General Relativity predicts that a circular binary with orbital frequency $f/2$ will emit GW with frequency $f$ and amplitude $\propto$$f^{2/3}$ \cite{Abbot17}, and when SMBH binaries have an orbital separation of $\sim$0.01\,parsecs (pc, $\sim$2000\,astronomical units), the orbits decay primarily through GW emission.  Because of this link between GW frequency and amplitude, the superposition of GWs from many SMBH binaries throughout the Universe is predicted to build up a GW background (GWB) with a characteristic GW strain $h_c$ following a power law in frequency \cite{Sesana04},

\begin{equation}
h_c(f) = \agwb \left(\frac{f}{\mathrm{yr}^{-1}}\right)^{\alpha}.
\end{equation}
The spectral index $\alpha$ is predicted to be $-2/3$ for GW-driven binary inspirals, while the dimensionless strain amplitude $\agwb$ incorporates the growth, masses, and merger rates of SMBHs. If SMBHs do not rapidly migrate to the centers of newly-merged galaxies, there will be relatively fewer wide binaries, reducing the GW power at low frequencies. Thus the measured GWB is expected to carry information about the distribution of SMBH masses and the dynamical evolution of SMBH binary systems \cite{Burke-Spolaor19}.

This GWB can be detected with ensembles of MSPs---known as pulsar timing arrays (PTAs) \cite{Sazhin78,Detweiler79}---by monitoring the times of arrival (TOAs) of the steady pulses from each pulsar, which arrive earlier or later than expected due to the spacetime perturbations.  Because the GWB is expected to be a sum of many individual sources, the induced TOA variations are random and differ for each pulsar, but have a common spectrum of power spectral densities, $P(f)$:
\begin{equation}
\label{eq:gwb_amp}
P(f) = \frac{\agwb^2}{12\pi^2}\left(\frac{f}{\mathrm{yr}^{-1}}\right)^{-\Gamma}\,\mathrm{yr}^{-3}, 
\end{equation}
with spectral index $\Gamma=3-2\alpha=13/3$ for SMBHs \cite{Sesana04}.  This functional form has more power at low frequencies so is referred to as a red spectrum.  For observations taken at an approximately fixed location (Earth), the GWB is expected to produce a signature quadrupolar pattern of TOA variations, known as the Hellings-Downs correlation \cite{Hellings83}.

\begin{figure}
\centering
\includegraphics[angle=0,width=0.98\linewidth]{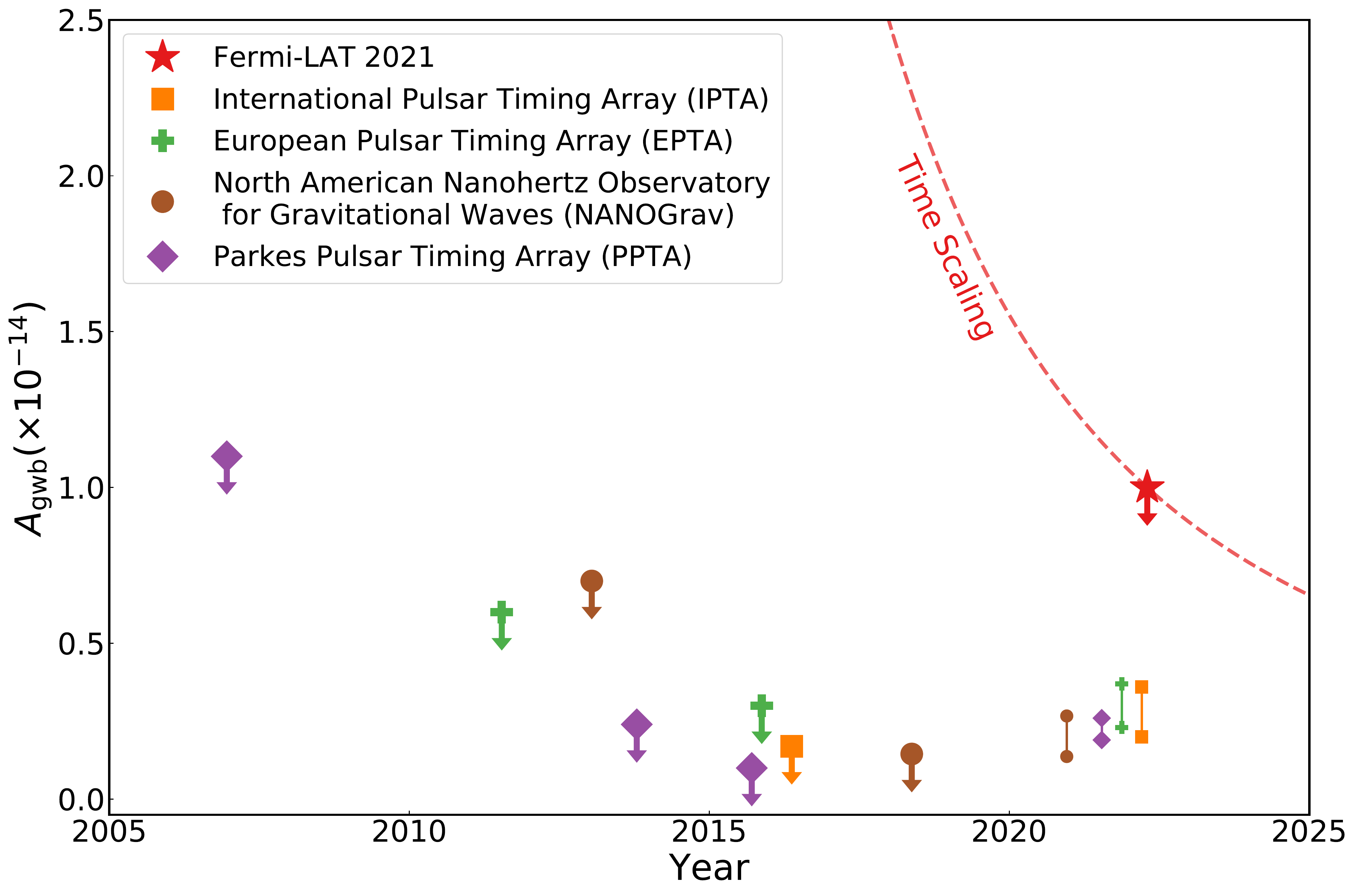}
\caption{\label{fig:limit_comp}
\baselineskip24pt
\textbf{Constraints on the gravitational wave background from radio and gamma-ray PTAs.} The inferred constraints on the GWB amplitude at 1\,yr$^{-1}$ ($\agwb$, data sources in Table S7) are plotted as a function of publication date and assume $\alpha=-2/3$, as predicted for the superposition of gravitational waves from merging black holes.  Colored symbols, indicated in the legend, correspond to each of the PTAs.  Upper limits at 95\% confidence are shown with arrows while amplitude ranges indicate detections of a common noise process.  The Fermi-LAT 95\% upper limit, $1.0\times10^{-14}$, uses data obtained through January, 2021 and is shown as a red star at an approximate publication date April, 2022.  The dashed red line indicates the expected scaling as the limit as a function of time.}
\end{figure}

Because the expected quadrupolar correlations are only about 10\% of the total signal, the GWB is predicted to first appear as a set of independent signals from each pulsar whose power spectra are all consistent with Equation \ref{eq:gwb_amp}, with the quadrupolar distribution only becoming evident in more sensitive observations.  Radio PTAs have reported a red spectrum process with modest statistical significance \cite{Arzoumanian20,Goncharov21b,Chen21,Antoniadis22}, but no Hellings-Downs correlation has been found.  These results could be compatible with $\alpha=-2/3$ and $\agwb\sim2-3\times10^{-15}$ at 1\,yr$^{-1}$; see Figure \ref{fig:limit_comp}. This would be consistent with some predictions for the GWB \cite{Burke-Spolaor19}, but because no spatial correlations have been detected, it could have other origins.

A potential alternative explanation for this signal is spin noise, approximately power-law red noise intrinsic to each pulsar, with some MSPs observed to have a spin noise spectral index ($\Gamma$) of 2--7 \cite{Alam21,Goncharov21a}.  Possible physical origins for spin noise include turbulence in the neutron star interior \cite{Melatos14} and systematic variations in the magnetic field and co-rotating plasma which govern the rotational energy loss of the pulsar \cite{Lyne10}.   Pulsars that have spin noise spectra with similar shapes but different amplitudes---inconsistent with a GWB---could masquerade as a common mode signal without a Hellings-Downs correlation \cite{Goncharov21b}.

Another potential noise source for radio pulsar timing arrays is the frequency-dependent effect of radio propagation through plasma, including the solar wind and the ionized interstellar medium (IISM).  Pulsed radio emission at frequency $\nu$ is delayed by time $\tau_{\mathrm{DM}}$
\begin{equation}
\tau_{\mathrm{DM}}=4.15\,\mathrm{ms}\times\left(\frac{\mathrm{DM\,cm^{3}}}{\mathrm{pc}}\right)\times\left(\frac{\nu}{\mathrm{GHz}}\right)^{-2},
\end{equation}
where DM is the dispersion measure, equal to the total electron column density.  The DM of a pulsar can vary with time, due to the relative motions of Earth and the pulsar.  Correcting for this effect requires repeated measurements using multi-frequency radio observations and the introduction of many additional degrees of freedom to timing models.  Because the propagation paths of radio waves through the IISM depend on $\nu$, the DM itself is frequency-dependent \cite{Cordes16,Donner20}, so some of this delay is intrinsically unmeasurable.  Other propagation effects include a broadening of the pulse that can only be corrected for bright pulsars, with some components also being unmeasurable \cite{Shannon17}.  Because the IISM is turbulent, these uncorrected delays introduce additional red noise to radio pulsar timing data.  The variable solar wind introduces similar dispersive delays which can in principle be measured like DM variations but are only partially included in current models \cite{Tiburzi21}.  Due to the wide angular extent of the solar wind, uncorrected delays would be correlated amongst pulsars. As with spin noise, IISM-induced noise with similar spectra could mimic a GWB signal.  Predicted noise amplitudes are similar to the expected GWB signal, but these predictions rely on assumptions about the turbulent spectra of the IISM which are poorly constrained by data \cite{Shannon17}.  See \cite{SOM} for further discussion of the modeling and impact of noise. 

Gamma-ray observations offer a potentially complementary approach: the much higher photon frequency means that the effects of the IISM and solar wind are negligible. The Large Area Telescope (LAT) \cite{Atwood09}, on the Fermi Gamma-ray Space Telescope, 
is sensitive to GeV gamma-ray photons emitted by MSPs.
Its 2.4 steradian field-of-view performs a continuous survey, covering the full sky every 2 orbits ($\sim$3\,hr). 
Its GPS clock records photon arrival times with $<$300\,ns precision \cite{Ajello21_onorbit}, enabling pulsar timing.
Analyses of the LAT survey data have detected 127 \cite{SOM, Smith19} of the over 400 known MSPs in the Milky Way.  The large MSP sample, long observing span, and instrumental stability enable a gamma-ray pulsar timing array whose characterization of spin noise and a potential GWB signal is free from IISM effects.

Using the 35 brightest and most stable $\gamma$-ray MSPs and 12.5\,yr of Fermi-LAT data, we searched for the GWB using two different techniques \cite{SOM}.
First, we implemented a coherent photon-by-photon analysis which retains $<$1$\,\mu$s resolution.  Second, for analysis with the established software used for radio data analysis, we directly measured TOAs from the LAT data \cite{Kerr15c}.  Because the TOA estimation procedure requires averaging up to one year of data, this method loses sensitivity to shorter-timescale signals, and only 29 of the 35 pulsars are suitable.

\begin{figure}
\centering
\includegraphics[angle=0,width=0.98\linewidth]{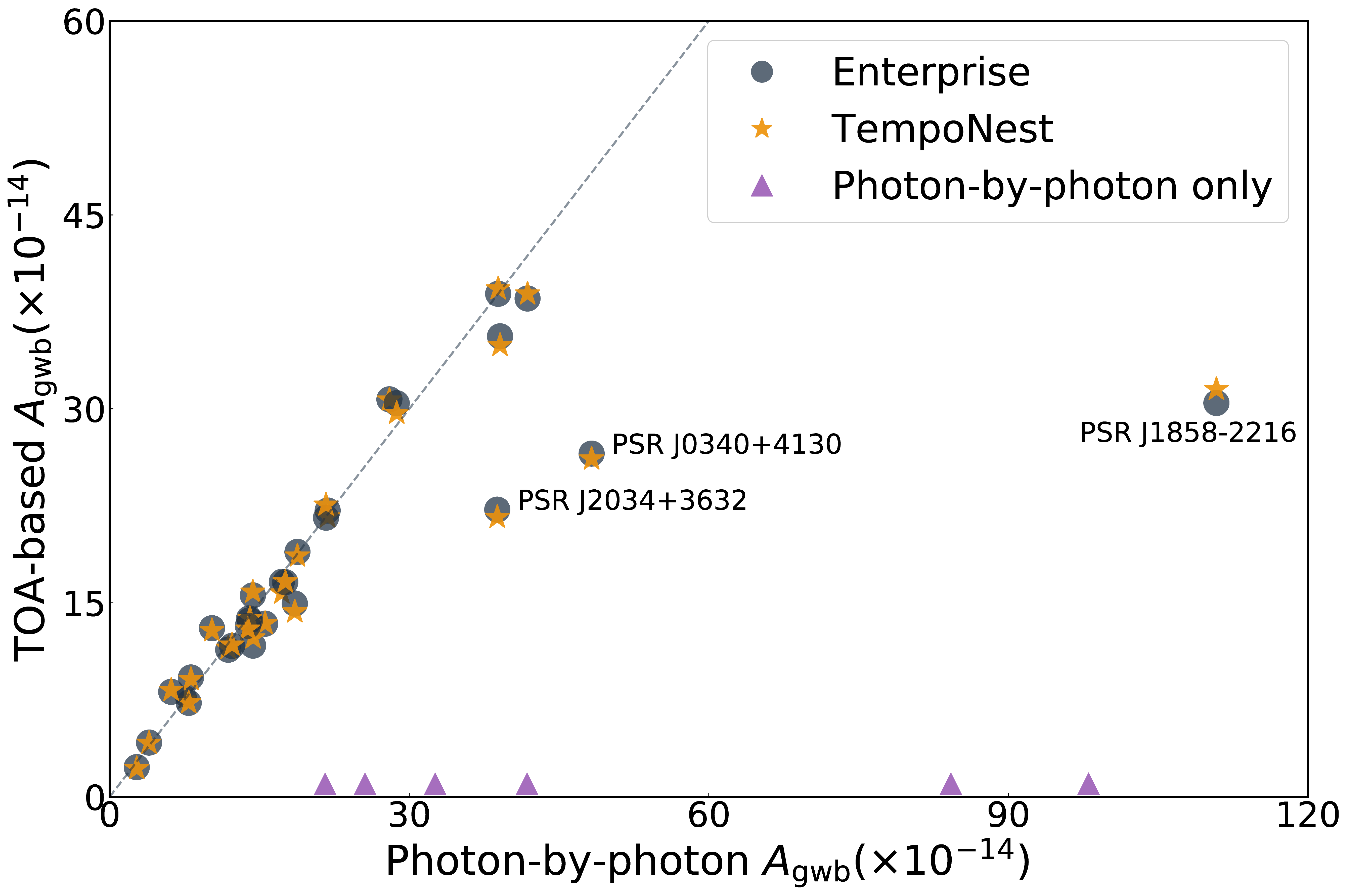}
\caption{
\baselineskip24pt
\label{fig:single_pulsar}\textbf{Comparison between $A_{\mathrm{gwb}}$ measurements from each pulsar using three analysis methods.}  Data points indicate the limits on an $\alpha=-2/3$ GWB for
35 MSPs computed with three methods: two TOA-based codes
\textsc{TempoNest} (orange stars) and \textsc{Enterprise} (blue circles) are shown as a function
of the limit from a photon-by-photon analysis (x axis).  The dashed line indicates equality between the results of the TOA-based and photon-by-photon methods.  Six pulsars (purple triangles) have only a photon-based analysis so are plotted arbitrarily at zero on the y axis.  The three labeled pulsars are outliers \cite{SOM}.}
\end{figure}

For each pulsar, we searched for spin noise and derived an upper limit on $\agwb$ using the photon-by-photon method and with two TOA-based software packages, 
\textsc{TempoNest} \cite{Lentati14} and \textsc{Enterprise}
\cite{enterprise}.  None of the pulsars show evidence for spin noise \cite{SOM} and the three different methods provide consistent results for each pulsar (Figure \ref{fig:single_pulsar}) except for three cases \cite{SOM}.

Three of the pulsars in our sample have spin noise measurements from radio PTAs.  Using the power spectral indices $\Gamma$ measured from the radio timing data, we calculated 95\% upper limits on spin noise amplitudes from the gamma-ray data.  Our limits are below the previously measured values for PSR~J0030$+$0451 (10\% of the measured value) and PSR~J1939$+$2134 (60--70\%) but are unconstraining for PSR~J0613$-$0200.  This discrepancy might indicate contamination by residual IISM effects on the radio-based spin noise and GWB signal measurements.

We combined the single pulsars into a pulsar timing array and
estimated $\agwb$ limits under a variety of scenarios, including marginalization over possible spin noise and uncertainties in the position of Earth relative to the Solar System barycenter, and both excluding and including the quadrupolar spatial correlations expected under General Relativity \cite{SOM}. The resulting representative 95\% confidence limit is $\agwb<1.0\times10^{-14}$ (Figure \ref{fig:limit_comp}), a factor of 3--5 higher than the red spectrum process detected by radio PTAs.

For an idealized PTA, when a potential GWB signal is weak compared to other noise, the signal-to-noise ratio grows proportionally to $\agwb^2\times t_{\mathrm{obs}}^\Gamma$ \cite{Siemens13,Pol21}, with $t_{\mathrm{obs}}$ the observing time span and $\Gamma=13/3$ for SMBHs as in Equation 2.  This means that upper limits on $\agwb$ improve following the relation $\agwb\propto t_{\mathrm{obs}}^{-13/6}$.
On the other hand, if the signal that terrestrial PTAs are currently detecting does arise from the GWB, then these PTAs are now in the
strong signal regime and their sensitivity will improve slowly ($\propto t_{\mathrm{obs}}^{-1/2}$).
The differing time scalings and noise sources allow the gamma-ray PTA data to distinguish residual IISM variations from a potential GWB signal.

The Fermi PTA data have an essentially constant experimental setup: the data are almost uninterrupted and calibrations have been constant for the full 12.5 year dataset.  Gamma-ray data are potentially less subject to astrophysical effects such as changes in the radio pulse shape \cite{SOM}.  This stability is particularly useful for probing GWs with frequencies below 0.1\,yr$^{-1}$.  Such low frequencies are predicted to constrain the spectral shape of the GWB which contains information about the physical sources \cite{Burke-Spolaor19}.
 
\begin{figure}
\centering
\includegraphics[angle=0,width=0.98\linewidth]{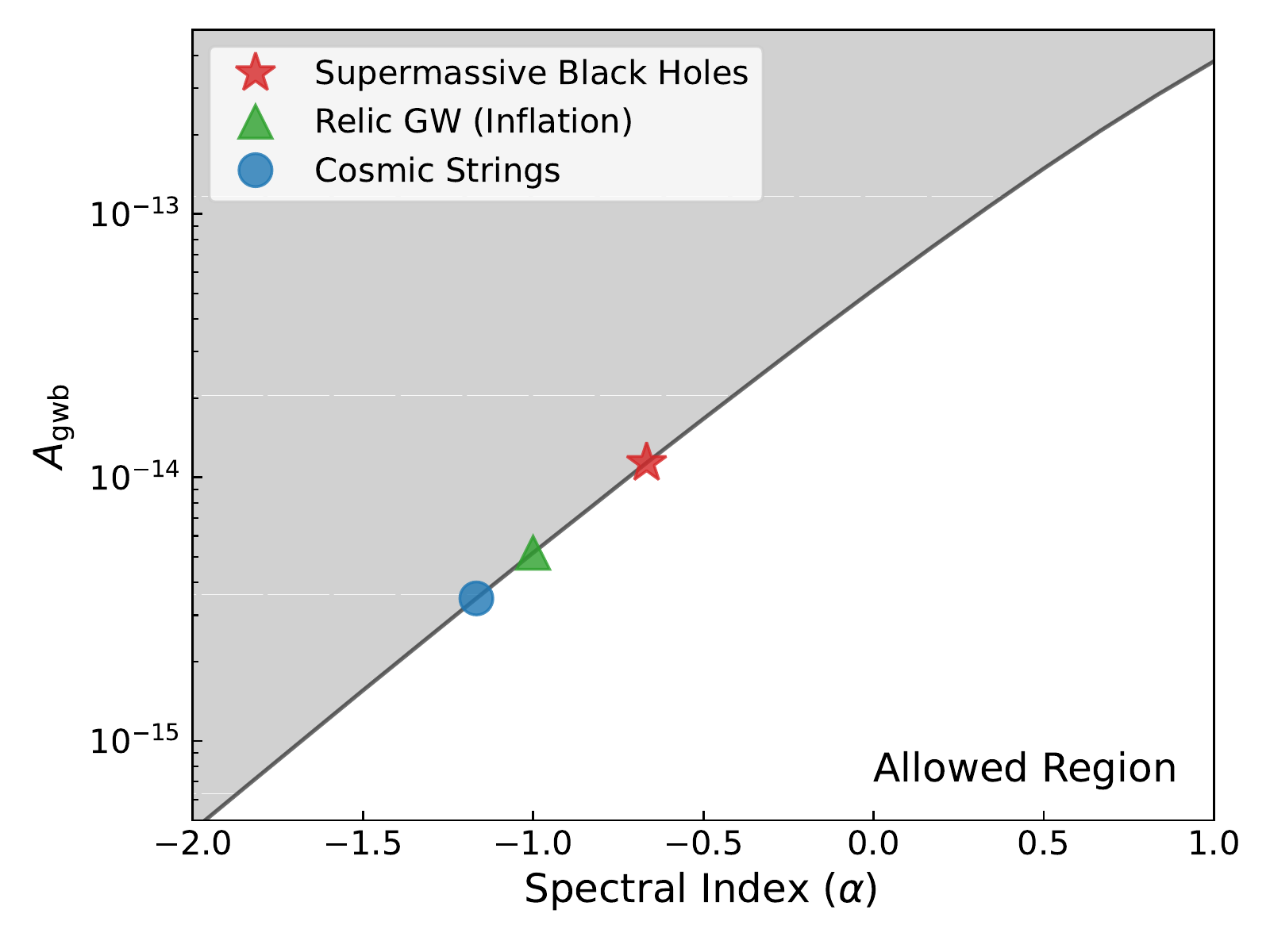}
\caption{
\baselineskip24pt
\label{fig:agwb_vs_indx}\textbf{Gamma-ray constraints on different types of GWB sources.} GWB amplitudes $\agwb$ for assumed spectral indices $\alpha$ in the shaded region are excluded with 95\% confidence.  The symbols indicate the values of $\alpha$ expected for SMBH binaries (red star, our fiducial result), gravitational waves generated during cosmic inflation (green triangle) and from hypothetical cosmic strings (blue circle).
}
\end{figure}

There are other potential sources of power-law GWBs with different spectral indices, $\alpha$, such as $\alpha=-1$ for relic GWs originating during scale-invariant inflation in the early Universe \cite{Zhao11}.  Decay of (hypothetical) cosmic strings could also produce power-law spectra under a variety of scenarios \cite{Damour05}. To constrain such sources, we computed corresponding 95\% upper limits on $\agwb$ at different values of $\alpha$ (Figure \ref{fig:agwb_vs_indx}).  Yet other models are not well described by power laws, but predict the largest signal in or near the PTA band \cite{Arzoumanian21_phase_transitions,Khmelnitsky14}.

To summarize, we have used the Fermi LAT dataset to construct a gamma-ray PTA.  This provides an independent method to search for signals detected by radio PTAs; unlike the radio PTAs, it is free from the effects of the IISM.
 Most of the pulsars are amenable to the TOA-based approach, and the resulting datasets are small compared to those of radio PTAs, enabling analysis alongside radio PTA data with little additional computational burden.

\bibliographystyle{Science}
\bibliography{sr.bib}

\noindent\textbf{Acknowledgements} We dedicate this work to the memory of our colleague, Jing Luo.  The authors are grateful to the insightful anonymous reviewers and to David Champion for an early review.  The Fermi-LAT Collaboration acknowledges support for LAT development, operation and data analysis from NASA and DOE (United States), CEA/Irfu and IN2P3/CNRS (France), ASI and INFN (Italy), MEXT, KEK, and JAXA (Japan), and the K.A.~Wallenberg Foundation, the Swedish Research Council and the National Space Board (Sweden). Science analysis support in the operations phase from INAF (Italy) and CNES (France) is also gratefully acknowledged. 
The National Radio Astronomy Observatory is a facility of the National Science Foundation operated under cooperative agreement by Associated Universities, Inc.  Pulsar research at UBC is supported by an NSERC Discovery Grant and by CIFAR.  Work at NRL is supported by NASA.
\textbf{Funding:} This work performed in part under DOE Contract DE-AC02-76SF00515. MK is supported by NASA grant NNG21OB03A.  ECF and NM are supported by NASA under award number 80GSFC21M0002.  TC is supported by NASA through the NASA Hubble Fellowship Program grant \#HST-HF2-51453.001.  KC is supported by a UBC Four Year Fellowship (6456).  SMR is a CIFAR Fellow and is supported by the NSF Physics Frontiers Center award 1430284.  The work of MASC and VG was supported by the grants PGC2018-095161-B-I00 and CEX2020-001007-S, both funded by MCIN/AEI/10.13039/501100011033 and by ERDF. VG has been supported by Juan de la Cierva-Incorporaci\'on IJC2019-040315-I grants. GZ acknowledge the financial support from the Slovenian Research Agency (grants P1-0031, I0-0033 and J1-1700). CJC acknowledges support from the ERC under the European Union’s Horizon 2020 research and innovation programme (grant agreement No. 715051; Spiders). SJS holds an NRC Research Associateship award at NRL.
\textbf{Author contributions:} MK (Kerr) conceived the project, implemented the gamma-ray analysis, and co-wrote the manuscript.  AP implemented the TOA-based analysis and co-wrote the manuscript.  DAS, PSR, MPR, and MK (Kramer) internally reviewed the manuscript.  BB, IC, HC, KC, LG, MJK, SMR, JR, RS, IS, ST, and GT contributed radio timing solutions.  Other co-authors acquired Fermi-LAT data and reviewed and contributed to the manuscript.  \textbf{Competing interests:} The authors declare they have no competing interests.  \textbf{Data and materials availability:} All raw data, processed data, reduced data, pulsar timing solutions, and software developed for this work are available in our data release \cite{zenodo}.

\baselineskip24pt


\section*{Supplementary materials}
Authors and Affiliations\\
Materials and Methods\\
Figs. S1 to S3\\
Tables S1 to S8\\
References \textit{(35-92)}

\clearpage




\topmargin 0.0cm
\oddsidemargin 0.2cm
\textwidth 16cm 
\textheight 21cm
\footskip 1.0cm



\baselineskip16pt

\newcommand{\hbAppendixPrefix}{S}
\renewcommand{\thefigure}{\hbAppendixPrefix\arabic{figure}}
\setcounter{figure}{0}
\renewcommand{\thetable}{\hbAppendixPrefix\arabic{table}} 
\setcounter{table}{0}
\renewcommand{\theequation}{\hbAppendixPrefix\arabic{equation}} 
\setcounter{equation}{0}


\newcommand{\eqgwbamp}{2}
\newcommand{\figsinglepulsar}{2}
\newcommand{\figagwbvsindx}{3}
\newcommand{\figlimitcomp}{1}

\begin{center}
{\large Supplementary Materials for}\\
\vspace{0.6cm}
{\bf A Gamma-ray Pulsar Timing Array Constrains the Nanohertz Gravitational Wave Background}\\

\vspace{0.5cm}
The Fermi-LAT Collaboration\\
\vspace{0.5cm}
Correspondence to: matthew.kerr@gmail.com, adityapartha3112@gmail.com

\end{center}

\vspace{0.5cm}
\noindent{\bf This PDF file includes:}\\

\indent Authors and Affiliations\\
\indent Materials and Methods\\
\indent Figs. S1 to S3\\
\indent Tables S1 to S8\\


\clearpage

\section*{Authors and Affiliations}
M.~Ajello$^{1}$, 
W.~B.~Atwood$^{2}$, 
L.~Baldini$^{3}$, 
J.~Ballet$^{4}$, 
G.~Barbiellini$^{5,6}$, 
D.~Bastieri$^{7,8}$, 
R.~Bellazzini$^{9}$, 
A.~Berretta$^{10}$, 
B.~Bhattacharyya$^{11}$, 
E.~Bissaldi$^{12,13}$
% ...
\\
\\
\begin{enumerate}
\item[1.] Department of Physics and Astronomy, Clemson University, Kinard Lab of Physics, Clemson, SC 29634-0978, USA
\item[2.] Santa Cruz Institute for Particle Physics, Department of Physics and Department of Astronomy and Astrophysics, University of California at Santa Cruz, Santa Cruz, CA 95064, USA
\item[3.] Universit\`a di Pisa and Istituto Nazionale di Fisica Nucleare, Sezione di Pisa I-56127 Pisa, Italy
\item[4.]  AIM, CEA, CNRS, Universit\'e Paris-Saclay, Universit\'e de Paris, F-91191 Gif-sur-Yvette, France
\item[5.] Istituto Nazionale di Fisica Nucleare, Sezione di Trieste, I-34127 Trieste, Italy
\item[6.]  Dipartimento di Fisica, Universit\`a di Trieste, I-34127 Trieste, Italy
\item[7.]  Istituto Nazionale di Fisica Nucleare, Sezione di Padova, I-35131 Padova, Italy
\item[8.]  Dipartimento di Fisica e Astronomia ``G. Galilei'', Universit\`a di Padova, I-35131 Padova, Italy 
\item[9.]  Istituto Nazionale di Fisica Nucleare, Sezione di Pisa, I-56127 Pisa, Italy 
\item[10.]  Dipartimento di Fisica, Universit\`a degli Studi di Perugia, I-06123 Perugia, Italy
\item[11.]  National Centre for Radio Astrophysics, Tata Institute of Fundamental Research, Pune 411 007, India
\item[12.]  Dipartimento di Fisica ``M. Merlin" dell'Universit\`a e del Politecnico di Bari, via Amendola 173, I-70126 Bari, Italy
\item[13.]  Istituto Nazionale di Fisica Nucleare, Sezione di Bari, I-70126 Bari, Italy
\item[14.]  W. W. Hansen Experimental Physics Laboratory, Kavli Institute for Particle Astrophysics and Cosmology, Department of Physics and SLAC National Accelerator Laboratory, Stanford University, Stanford, CA 94305, USA
\item[15.]  Istituto Nazionale di Fisica Nucleare, Sezione di Torino, I-10125 Torino, Italy 
\item[16.]  Dipartimento di Fisica, Universit\`a degli Studi di Torino, I-10125 Torino, Italy
\item[17.]  Laboratoire Leprince-Ringuet, \'Ecole polytechnique, Centre national de la recherche scientifique / Institut national de physique nucl\'eaire de physique des particules, F-91128 Palaiseau, France
\item[18.]  Deutsches Elektronen Synchrotron, D-15738 Zeuthen, Germany 
\item[19.]  Department of Physics and Astronomy, Louisiana State University, Baton Rouge, LA 70803, USA
\item[20.]  Institut f\"ur Theoretische Physik and Astrophysik, Universit\"at W\"urzburg, D-97074 W\"urzburg, Germany
\item[21.]  Istituto Nazionale di Astrofisica-Istituto di Astrofisica Spaziale e Fisica Cosmica Milano, via E. Bassini 15, I-20133 Milano, Italy
\item[22.]  Agenzia Spaziale Italiana, Via del Politecnico, snc, I-00133 Roma, Italy 
\item[23.]  Istituto Nazionale di Fisica Nucleare, Sezione di Roma ``Tor Vergata", I-00133 Roma, Italy 
\item[24.]  Space Science Data Center, Agenzia Spaziale Italiana, Via del Politecnico, snc, I-00133, Roma, Italy 
\item[25.]  Jodrell Bank Centre for Astrophysics, Department of Physics and Astronomy, The University of Manchester, M13 9PL, UK 
\item[26.]  Albert-Einstein-Institut, Max-Planck-Institut f\"ur Gravitationsphysik, D-30167 Hannover, Germany 
\item[27.]  Leibniz Universit\"at Hannover, D-30167 Hannover, Germany 
\item[28.]  Laboratoire de Physique et Chimie de l'Environnement et de l'Espace, Universit\'e d'Orl\'eans / Centre national de la recherche scientifique, F-45071 Orl\'eans Cedex 02, France 
\item[29.]  Station de radioastronomie de Nan\c{c}ay, Observatoire de Paris, Centre national de la recherche scientifique/Institut national des sciences de l'Univers, F-18330 Nan\c{c}ay, France 
\item[30.] Instituto de F\'isica Te\'orica (Consejo Superior de Investigaciones Científicas), Universidad Aut\'onoma de Madrid, E-28049 Madrid, Spain
\item[31.]  Departamento de F\'isica Te\'orica, Universidad Aut\'onoma de Madrid, 28049 Madrid, Spain 
\item[32.]  Department of Astronomy, University of Maryland, College Park, MD 20742, USA 
\item[33.]  Cornell Center for Astrophysics and Planetary Science and Department of Astronomy, Cornell University, Ithaca, NY 14853, USA 
\item[34.]  Department of Physics and Astronomy, University of British Columbia, 6224 Agricultural Road, Vancouver, BC V6T 1Z1, Canada 
\item[35.]  Istituto Nazionale di Fisica Nucleare, Sezione di Perugia, I-06123 Perugia, Italy 
\item[36.]  Istituto Nazionale di Astrofisica-Istituto di Radioastronomia, I-40129 Bologna, Italy 
\item[37.]  Dipartimento di Matematica e Fisica ``E. De Giorgi", Universit\`a del Salento, Lecce, Italy 
\item[38.]  Istituto Nazionale di Fisica Nucleare, Sezione di Lecce, I-73100 Lecce, Italy 
\item[39.]  Laboratoire d'Annecy-le-Vieux de Physique des Particules, Universit\'e de Savoie, Centre national de la recherche scientifique/Institut national de physique nucl\'eaire de physique des particules, F-74941 Annecy-le-Vieux, France 
\item[40.]  Grupo de Altas Energ\'ias, Universidad Complutense de Madrid, E-28040 Madrid, Spain 
\item[41.]  NASA Goddard Space Flight Center, Greenbelt, MD 20771, USA 
\item[42.]  Center for Research and Exploration in Space Science and Technology (CRESST), Greenbelt, MD 20771, USA 
\item[43.]  Ruhr University Bochum, Faculty of Physics and Astronomy, Astronomical Institute, 44780 Bochum, Germany 
\item[44.]  Department of Physical Sciences, Hiroshima University, Higashi-Hiroshima, Hiroshima 739-8526, Japan 
\item[45.]  Friedrich-Alexander Universit\"at Erlangen-N\"urnberg, Erlangen Centre for Astroparticle Physics, Erwin-Rommel-Str. 1, 91058 Erlangen, Germany 
\item[46.]  Max-Planck-Institut f\"ur Physik, D-80805 M\"unchen, Germany 
\item[47.]  The George Washington University, Department of Physics, 725 21st St, NW, Washington, DC 20052, USA 
\item[48.]  Georg-August University G\"ottingen, Institute for theoretical Physics - Faculty of Physics, Friedrich-Hund-Platz 1, D-37077 G\"ottingen, Germany 
\item[49.]  Los Alamos National Laboratory, Los Alamos, NM 87545, USA 
\item[50.]  University of North Florida, Department of Physics, 1 UNF Drive, Jacksonville, FL 32224, USA 
\item[51.]  Yunnan Observatories, Chinese Academy of Sciences, 396 Yangfangwang, Guandu District, Kunming 650216, P.~R.~China  
\item[52.]  Key Laboratory for the Structure and Evolution of Celestial Objects, Chinese Academy of Sciences, 396 Yangfangwang, Guandu District, Kunming 650216, P.~R.~China  
\item[53.]  Science Institute, University of Iceland, IS-107 Reykjavik, Iceland 
\item[54.]  Nordita, Royal Institute of Technology and Stockholm University, Roslagstullsbacken 23, SE-106 91 Stockholm, Sweden 
\item[55.]  Space Science Division, Naval Research Laboratory, Washington, DC 20375-5352, USA 
\item[56.]  Max-Planck-Institut f\"ur Radioastronomie, Auf dem H\"ugel 69, D-53121 Bonn, Germany 
\item[57.]  University of Manchester, Manchester, M13 9PL, UK 
\item[58.]  Department of Physics, Kungliga Tekniska h{\"o}gskolan Royal Institute of Technology, AlbaNova, SE-106 91 Stockholm, Sweden 
\item[59.]  The Oskar Klein Centre for Cosmoparticle Physics, AlbaNova, SE-106 91 Stockholm, Sweden
\item[60.]  School of Education, Health and Social Studies, Natural Science, Dalarna University, SE-791 88 Falun, Sweden 
\item[61.] CAS Key Laboratory for Research in Galaxies and Cosmology, Department of Astronomy, University of Science and Technology of China, Hefei 230026, P.~R.~China
\item[62.]  Department of Astronomy, School of Physical Sciences, University of Science and Technology of China, Hefei, Anhui 230026, P.~R.~China

\item[63.]  Institut f\"ur Astro- und Teilchenphysik, Leopold-Franzens-Universit\"at Innsbruck, A-6020 Innsbruck, Austria 
\item[64.]  Department of Physics and Center for Space Sciences and Technology, University of Maryland Baltimore County, Baltimore, MD 21250, USA 
\item[65.]  Department of Physics, Faculty of Science, Mahidol University, Bangkok 10400, Thailand 
\item[66.]  Hiroshima Astrophysical Science Center, Hiroshima University, Higashi-Hiroshima, Hiroshima 739-8526, Japan 
\item[67.]  Istituto Nazionale di Fisica Nucleare, Sezione di Trieste, and Universit\`a di Trieste, I-34127 Trieste, Italy 
\item[68.]  Department of Physics and Astronomy, University of Denver, Denver, CO 80208, USA 
\item[69.]  Osservatorio Astronomico di Trieste, Istituto Nazionale di Astrofisica, I-34143 Trieste, Italy 
\item[70.]  Center for Space Studies and Activities ``G. Colombo", University of Padova, Via Venezia 15, I-35131 Padova, Italy 
\item[71.]  Korea Astronomy and Space Science Institute, 776 Daedeokdae-ro, Yuseong-gu, Daejeon 30455, Korea 
\item[72.]  Department of Physics, American University, Washington, DC 20016, USA 
\item[73.]  National Radio Astronomy Observatory, Charlottesville, VA 22903, USA 
\item[74.]  Centre for Astro-Particle Physics, University of Johannesburg, PO Box 524, Auckland Park 2006, South Africa 
\item[75.]  Department of Physics, University of Johannesburg, PO Box 524, Auckland Park 2006, South Africa 
\item[76.]  Department of Physics, The University of Hong Kong, Pokfulam Road, Hong Kong, P.~R.~China  
\item[77.]  Laboratory for Space Research, The University of Hong Kong, Hong Kong, China 
\item[78.]  Astrobiology and Space Sciences Division, NASA Ames Research Center, Moffett Field, CA 94035-1000,
USA (retired)
\item[79.]  Laboratoire Univers et Particules de Montpellier, Universit\'e de Montpellier, CNRS/IN2P3, F-34095 Montpellier, France 
\item[80.]  NYCB Real-Time Computing Inc., Lattingtown, NY 11560-1025, USA 
\item[81.]  Centre d'\'Etudes Nucl\'eaires de Bordeaux Gradignan, IN2P3/CNRS, Universit\'e Bordeaux 1, BP120, F-33175 Gradignan Cedex, France 
\item[82.]  Laboratoire d'Astrophysique de Bordeaux, Universit\'e de Bordeaux, CNRS, B18N, all\'ee Geoffroy Saint-Hilaire, F-33615 Pessac, France 
\item[83.]  Australian Research Council Centre of Excellence for Gravitational Wave Discovery (OzGrav), Centre for Astrophysics and Supercomputing, Mail H29, Swinburne University of Technology, PO Box 218, Hawthorn, VIC 3122, Australia 
\item[84.]  Centre for Astrophysics and Supercomputing, Swinburne University of Technology, PO Box 218, Hawthorn Victoria 3122, Australia 
\item[85.]  Purdue University Northwest, Hammond, IN 46323, USA 
\item[86.]  Resident at Naval Research Laboratory, Washington, DC 20375, USA 
\item[87.]  New York University Abu Dhabi, P.O. Box 129188, Abu Dhabi, United Arab Emirates 
\item[88.]  Department of Physics and Astronomy, West Virginia University, Morgantown, WV 26506-6315, USA 
\item[89.]  Instituci\'o Catalana de Recerca i Estudis Avan\c{c}ats (ICREA), E-08010 Barcelona, Spain
\item[90.]  Institute of Space Sciences (ICE, CSIC), Campus UAB, Carrer de Magrans s/n, E-08193 Barcelona, Spain
\item[91.]  Institut d'Estudis Espacials de Catalunya (IEEC), 08034 Barcelona, Spain
\item[92.]  Praxis Inc., Alexandria, VA 22303, USA 
\item[93.]  Center for Astrophysics and Cosmology, University of Nova Gorica, Nova Gorica, Slovenia
\item[$\dagger$] matthew.kerr@gmail.com, adityapartha3112@gmail.com
\end{enumerate}

\section*{Materials and Methods}

\section*{Pulsar Timing using Radio and Gamma-ray Observations}
The spin phase at time $t$, $\phi(t)$, increases by 1 each time a pulsar rotates, and because the pulsation mechanism is fixed to the star, $\phi(t)$ can be inferred by observing pulses.  Pulsar timing is the measurement of these pulse arrival times (TOAs) and comparison with a timing model that predicts $\phi(t,\lambda)$.  The astrophysical parameters $\lambda$ leave a characteristic imprint on the timing residuals between data and model: a spin frequency error $\delta\nu$ produces linear residuals, $\delta\dot{\nu}$ leaves quadratic residuals, a position error induces an annual sinusoid, etc.~\cite{Lorimer04}.  The goal of pulsar timing is to measure $\lambda$ and thus characterize a wide range of astrophysical phenomena and constrain fundamental physics \cite{Taylor79,Kramer06b,Kramer09,Will14}.

During a typical $\sim$0.1--1\,hr observation with a radio telescope, light collected from the antenna is transduced, amplified, digitized, dedispersed, and filtered into frequency channels.  Pulses are stacked into a single pulse profile by folding the data at the instantaneous pulsar spin period.  The observation time is recorded using a precise clock, and the offset of the recorded pulse to this time yields the pulse TOA.  The uncertainty of a TOA can be estimated from the random noise in the pulse profile, which arises from electronics noise, background radiation from astrophysical sources, and terrestrial spillover into the antenna.  The resulting uncertainty is Gaussian, so a typical radio timing analysis optimizes the timing model parameters and assesses goodness-of-fit by minimizing $\chi^2(\lambda)$. 

The Fermi-LAT, on the other hand, collects individual gamma rays.  The arrival time of the $i$th photon, $t_i$, is recorded with $\sim$300\,ns precision, but aside from its energy, a photon carries no additional information: it could be from any pulsar phase $\phi$ or even from a background source, so these $t_i$ cannot be interpreted as TOAs.  In some cases, histograms in $\phi$ can be built up over a long enough time---hours for bright pulsars, years for the faintest---that a pulse profile emerges.  Analogously to folded radio profiles, comparing these histograms to the assumed template $f(\phi)$ can yield a TOA \cite{Ray11}, and $\chi^2(\lambda)$ methods may be applicable. (Radio and gamma-ray pulse profiles differ, but we use $f(\phi)$ for the pulse profile in the relevant band.)

For many of the faint MSPs used in this work, there is insufficient integration time to build up a gamma-ray pulse profile and estimate a TOA.  For instance, resolving the annual sinusoidal residuals from a position error requires sampling of at least 2 TOAs per year, and preferably faster to avoid a systematic error.  Instead, we can use a timing model to evaluate the phase $\phi(t_i,\lambda)$ at each individual photon time and then gauge the agreement of the resulting distribution with an assumed template $f(\phi)$ using the likelihood (Equation \ref{eq:logl_notn}).  Because the LAT has a broad, energy-dependent angular resolution, photons from different sources overlap and we must also account for the background, which we do by computing the probability weight $0<w_i<1$ that the $i$th photon originates from the pulsar \cite{Bickel08,Kerr11,Bruel19}.  Using a normalized ($\int_0^1 f(\phi)\,d\phi=1$) pulse profile model, the Poisson likelihood for the data, $\mathcal{L}_{\mathrm{data}}$, is
\begin{equation}
\label{eq:logl_notn}
\log \mathcal{L}_{\mathrm{data}}(\lambda)=\sum_i \log \bigg(w_i f(\phi[t_i,\lambda]) +
(1-w_i)\bigg).
\end{equation}
By maximizing $\mathcal{L}_{\mathrm{data}}(\lambda)$, we obtain optimal estimates for parameters $\lambda$ while preserving the $<$1\,\textmu{}s resolution of the LAT.

Unlike in the $\chi^2(\lambda)$ radio case, there are no residuals because there are no direct phase measurements (TOAs) with which to compare the model.  This makes it more challenging to assess the goodness of fit and select the most favored models \cite{Protassov02}.  Our GWB analysis (below) uses both TOA-based and photon-by-photon (likelihood) methods.  We view the latter as more fundamental, but the former allows consistency checks with existing methods used for radio PTAs.

\section*{Noise in Pulsar Timing Array Data}

PTAs \cite{Sazhin78,Detweiler79} extend pulsar timing to ensembles (arrays) of pulsars to search for correlated signals such as low-frequency gravitational waves.  These subtle signatures can be obscured by noise, and many noise sources must be accounted for in radio PTA data to reach a typical amplitude in the timing residuals of $<$100\,ns.  To place the Fermi PTA results in context, we give an overview of the radio PTA noise budget below, drawing distinctions to gamma-ray observations where appropriate.  These main noise sources are summarized in Table \ref{tab:noise_budget}, and in \cite{Verbiest18}.

\subsection*{White Noise}

White noise ($<$1\,day time scales) cannot produce GWB-like signals, but it must be modeled to estimate the precision of timing model parameters.  Most analyses use the parameters \texttt{EFAC} and \texttt{EQUAD} \cite{Verbiest16}, which modify the measurement uncertainty $\sigma_i$ of a TOA according to $\sigma_i^2 \rightarrow
\mathtt{EFAC}^2\sigma_i^2 + \mathtt{EQUAD}^2$.  Additional \texttt{ECORR} parameters can be used to describe correlated noise, e.g.~jitter.  Because the noise sources depend on observing frequency, pulsar, and observing system, tens to hundreds of such parameters may be needed to represent the excess white noise in a radio PTA data set. The mapping of these parameters onto the actual noise sources may be imperfect, leaving un- or over-modeled white noise in the residuals.

For any relevant PTA pulsar, gamma-ray measurement uncertainties are dramatically larger, and poor sensitivity is the major weakness of gamma-ray pulsar timing.

\begin{table}
\centering
\caption{\label{tab:noise_budget}\textbf{Sources of noise in radio and gamma-ray pulsar timing array data.} The list is incomplete and qualitative: for each noise source, we attempt to give an estimate of the importance (amplitude in residuals, or difficulty of complete mitigation) and complexity (degrees of freedom (d.o.f.) in existing models for a typical pulsar).  Dashes indicate that the entry is not applicable, and a question mark that its impact is unknown.}
\vspace{0.5cm}
\resizebox{\textwidth}{!}{
\begin{tabular}{l | l | r | l | r | l  }
\multicolumn{1}{l|}{} & \multicolumn{2}{c|}{Radio} & \multicolumn{2}{c|}{Gamma ray} & \multicolumn{1}{l}{}\\
\hline
\multicolumn{1}{c|}{Noise Source} & \multicolumn{1}{c|}{Impact} & \multicolumn{1}{c|}{d.o.f.} & \multicolumn{1}{c|}{Impact} & \multicolumn{1}{c|}{d.o.f.} & \multicolumn{1}{c|}{Note}\\
\hline
\hline
\multicolumn{1}{l}{White Noise} \\
\hline
Measurement & moderate   & --  & major & -- &  Sensitivity is major limiting factor for gamma rays.\\
RFI & minor & ? & -- & -- & RFI varies widely between observing systems.\\
Calibration & minor & ? & -- & -- & Affects certain pulsars/observing systems.\\
Jitter      & moderate & 10s & --     & -- & Jitter affects high signal-to-noise observations.\\
\hline
\multicolumn{1}{l}{Red Noise} \\
\hline
DM variation & major & 100s & -- & -- & DM(t) drives radio PTA observing strategies.\\			
Solar wind & moderate & $\sim$10s & -- & -- & Solar wind mitigation is poorly supported. \\
Scattering & moderate & 100s & -- & -- & Affects some pulsars/low radio frequencies. \\
Pulse variability & moderate & 0--10s & -- & ? & No gamma-ray MSP pulse profile changes known. \\
Discontinuities & moderate & 10s & -- & -- & LAT data are continuous, not a general property.\\
Spin noise & major & 10s--100s & major & 10s & Fewer d.o.f. needed for less precise LAT data.\\

\hline 
\end{tabular}
}
\end{table}

\paragraph{Jitter} 
Intrinsic pulse shape variations and the finite number of pulses received in an observation cause a TOA bias. This source of noise is called jitter \cite{Shannon14,Parthasarathy21}.
It requires a dedicated noise model because its impact depends on pulsar brightness, which varies due to IISM scintillation (below), and because it is correlated over the observing bandwidth \cite{Shannon14}.  Because $\ll$1 photon is received from each pulse on average, jitter is fundamental to gamma-ray observations and is accounted for in Poisson statistics. 

\paragraph{Radio-frequency interference (RFI)}  
RFI affects all radio observatories but is less severe for those in remote areas.  Strong RFI can render an observation unusable.  Fainter RFI can often be characterized as narrowband---contaminating portions of the observing band, intermittently or continuously---or impulsive---affecting much of the band briefly.  
RFI biases TOA measurements \cite{Kerr20}: impulsive RFI alters the observed pulse profile while narrowband RFI changes the effective observing frequency, inducing residuals from DM(t) corrections (below).  RFI noise is typically white, but a changing RFI environment can introduce biases on longer time scales. There is no analogous interference in Fermi-LAT data.

\paragraph{Polarization} 
Pulsars are highly polarized radio sources\cite{Dai15}, so recording a stable pulse profile requires accurate polarization calibration.  Since most radio telescopes use an altitude-azimuth mount, the measured pulsar polarization angle depends on the source elevation and must be corrected to obtain the intrinsic value.  If a radio receiver has non-zero cross-polarization, this alters the pulse profile \cite{vanStraten10}.  Using bright pulsars to characterize the receiver response as a function of parallactic angle \cite{VanStraten13} can eliminate much of the apparent TOA variation.  As with RFI, observational bias can couple this white noise to longer time scales. Gamma-ray pulsar polarization is unknown, but Fermi-LAT is very insensitive to photon polarization \cite{Giomi17}.


\subsection*{Red Noise}
Red noise operates on longer time scales and includes signals with power spectral densities similar to that of the GWB, so it must be modeled.

\paragraph{Spin Noise}
Spin noise, also called timing noise or intrinsic noise, is well known in young, slowly spinning pulsars \cite{Parthasarathy19}, but is also present in MSPs with an amplitude that is related to the pulsar spin-down power \cite{Shannon10}.  This intrinsic noise is generally adequately modeled with a power-law spectrum of the same form as Equation \eqgwbamp{}, and represented in data either as long-term correlations between the TOAs, with a non-diagonal covariance matrix \cite{vanHaasteren11,Coles11}, or using a Fourier expansion with the coefficients constrained to follow the assumed power spectrum \cite{Lentati13}. We adopt the latter approach. The prescriptions both assume that spin noise originates from a stationary process.  This assumption may be appropriate if such spin noise originates from superfluid turbulence within the neutron star \cite{Melatos14} or fluctuations within the pulsar magnetosphere, but it may be a poor approximation if such noise arises from e.g.~switches between metastable equilibria of the magnetosphere \cite{Kramer06,Lyne10,Hermsen13}.  Most proposed mechanisms predict identical spin noise in the radio and gamma-ray bands.

\paragraph{Effects of the Ionized Interstellar Medium}
The ionized interstellar medium (IISM) is turbulent \cite{Cordes85a} and contains structures that act as lenses \cite{Coles15}.  Radio waves from a pulsar encounter a continually changing electron density and bend proportionally to the inverse square of the observing radio frequency $\nu^{-2}$. The wide range of observational consequences includes dispersive delays, strong intensity variations (scintillation), broadening of pulses due to multipath scattering, and higher-order effects, such as apparent position shifts \cite{Coles10,Shannon17,Stinebring13}.

The line of sight through the IISM changes constantly due to the relative motion of Earth and the source. The most noticeable effect is scintillation: frequency- and time-dependent variations in the received intensity.  Scintillation can render some observations useless if the pulsar is too faint, and it shifts the effective observing frequency by enhancing or depressing the signal over the observational band, magnifying the effects of DM uncertainty discussed below.

The dominant effect of the IISM on residuals is the dispersive delay $\tau(t)\propto\,\mathrm{DM}(t)/\nu^2$, which changes measurably on timescales of days to weeks \cite{Keith13}.  Low-frequency measurements yield higher DM precision, and since their inception, PTAs have monitored DM(t) using multiple receivers covering a wide range of frequencies (e.g.~350\,MHz to 4\,GHz). Wider bandwidth receivers \cite{Hobbs20} and high-cadence observations by low-frequency observatories \cite{Amiri21, Donner20} also improve DM measurements. 
However, the apparent DM also depends on $\nu$ because lower-frequency radio waves scatter through a larger volume of the IISM, so even precise DM(t) measurements cannot fully correct $\tau(t)$ at other radio frequencies \cite{Cordes16, Donner19}.

DM variation contributes potentially hundreds of degrees of freedom to a pulsar timing model.  One approach uses \texttt{DMX} parameters \cite{Alam21} to tabulate DM measurements from TOAs within discrete time segments.  Another makes use of a constrained stochastic model for DM(t) (see spin noise above), which has fewer effective degrees of freedom but can only model stationary DM variations, which is generally insufficient to capture observed variations \cite{Keith13,Coles15,Lam18}.  
These different approaches to DM modeling can produce discrepant results: NANOGrav, using \texttt{DMX} \cite{Arzoumanian20}, and the PPTA, using the stochastic model \cite{Goncharov21a}, find marginally different values for spin noise for pulsars common to the two PTAs (see below).
 
 Other IISM effects can be approximated using corrections of the form $\nu^{n}$, e.g.~$\nu^{-4}$ for multipath scattering, and other powers for higher-order effects \cite{Shannon16}, though only $\nu^{-2}$ corrections are in widespread use. Scattering can be corrected precisely for a few bright MSPs using cyclic spectroscopy \cite{Demorest11}. A more universal method to incorporate such corrections is using a pulse portraiture, a model of a pulse profile over typically a factor of two or more of bandwidth \cite{Pennucci14}.  This portrait attempts to track the intrinsic variation in the pulse shape with frequency, and then is convolved with models of the IISM, e.g.~a $\nu^{-2}$ kernel to account for a changing DM, a $\nu^{-4}$ kernel for altered scattering, etc.

 In summary, the monitoring, measurement, and interpolation of IISM-induced residuals is an observationally and computationally expensive endeavor and a source of unmodeled error. Radio PTAs have managed to reduce these effects to the $\lessapprox$1\,\textmu{}s level. None of the IISM effects described above, or solar wind effects described below, affect gamma-ray timing, which is also immune to the eclipses and DM variations affecting compact interacting binaries \cite{Main18}.

\paragraph{The solar wind}
The solar wind makes a small contribution (typically $10^{-4}$--$10^{-3}$\,pc\,cm$^{-3}$) to the total DM for any pulsar, but it varies daily due to solar activity \cite{Tiburzi19, Tiburzi21} and annually, as the apparent pulsar-Sun angular separation changes.  Pulsars near the ecliptic, e.g.~PSR~J0030$+$0451, are the most strongly affected. Existing models and observational strategies are insufficient to capture the contribution of the solar wind to pulsar timing residuals \cite{Tiburzi19}, so most PTA data analyses either combine solar wind variations with DM variations or employ a static model.  Because the coherent structure of the solar wind subtends large portions of the sky, uncorrected solar wind variations can contribute correlated noise to pulsars \cite{Tiburzi16}.

\paragraph{Pulse profile variations}
Pulse profile variations induce apparent variations in TOAs and cannot be easily mitigated.  Observed variations in PSR~J1643$-$1224 \cite{Shannon16} and PSR~J0437$-$4715 \cite{Kerr20} have been attributed to a transient magnetospheric reconfiguration, while alterations in the pulse profile of PSR~J1713$+$0747 may be associated with rapid apparent DM variations and recovery \cite{Lin21}.  No pulse profile variations have been observed in gamma-ray MSP profiles, though correlated variations in intensity and pulse shape have been observed for the young PSR~J2021$+$4026 \cite{Allafort13}.  A stable gamma-ray pulse profile is expected due to the likely origin of pulsed gamma-ray emission is from synchro-curvature radiation precipitated by large-scale, stable electric fields in the outer magnetosphere or current sheet \cite{Kalapotharakos19}.

\paragraph{Discontinuities} 
Any observation of a pulsar is referenced to international time standards, a well-established process with typical errors expected to be $<$10\,ns.  However, instrumentation changes can alter the signal propagation time through the observing system, shifting the measured pulse phase relative to its true value.  These phase shifts must be measured from or modelled in the data, incurring potential systematic error or adding typically 10--20 degrees of freedom in timing models \cite{Kerr20}.  Fitting for these shifts acts as a high-pass filter, reducing sensitivity to low-frequency signals.  Although not a generic property of gamma-ray pulsar timing, the LAT has a stable GPS clock and the LAT itself has been operating almost continuously with nearly-constant instrumental properties \cite{Ajello21_onorbit}.  Any calibration errors are static, resulting in static biases but no time-domain noise.

\paragraph{Implications for gamma-ray analysis}
In summary, mitigating the noise sources for radio observations requires multi-frequency data, large fractional bandwidths, and homogeneous and regular monitoring, a substantial practical challenge.  In contrast, gamma-ray data only require spin noise and Poisson noise models.  This eases computational requirements and reduces systematic uncertainty. Due to continual all-sky monitoring, when a new MSP is discovered, archival LAT data can provide a full pulse timing history. The data span for each pulsar is uniform, ensuring that each pulsar is sensitive to the same spectrum of gravitational waves and enabling simple computational approaches. 

\section*{Data Preparation}
\label{sec:data}

\paragraph{Photon data}
Properties of each gamma ray were determined by reconstructing the particle interactions in the LAT detector into measured
quantities of incident direction, energy, and arrival time \cite{Atwood09}, with timestamping precision $<$300\,ns \cite{Ajello21_onorbit}.  We began with a set of 127 MSPs with gamma-ray counterparts in 4FGL-DR2, the second release \cite{4FGL_DR2} of the fourth Fermi-LAT gamma-ray source catalog \cite{4FGL}.  For each pulsar, we
downloaded data in the form of FT1 files (event lists) and FT2 files (tabulations of spacecraft position) from the Fermi Science Support
Center \cite{fssc_url} and
processed it using the Fermi Science Tools v2.0.0 \cite{fermitools}.     
We selected all data between 2008 Aug 04 (Modified Julian Day (MJD) 54682) and 2021 Jan 28
(MJD 59242)
with reconstructed energy between 0.1\,GeV and 10\,GeV, measured zenith
angle $<$100$^\circ$, and a reconstructed incidence direction placing
the photon within 3$^\circ$ of the pulsar position.  Using the 4FGL-DR2 sources models, we assigned each photon a weight (as in Equation \ref{eq:logl_notn}) using the gtsrcprob Science Tool and
restricted attention to events with $w_i>0.05$.  This selection is intended to retain the great majority of the pulsar signal while eliminating background photons that increase
computational costs.  We used the \textsc{PINT} software package \cite{Luo21} to evaluate timing models and assign spin phase to photons.  All raw and processed data, pulsar timing solutions, and software developed for this work are available in our data release \cite{zenodo}.  


\begin{table}
\footnotesize
\centering
\caption{\label{tab:ephemerides}\textbf{Properties of the 35 MSP ephemerides used for GWB
analyses.} Precise coordinates are provided in the data release \cite{zenodo}.  The observatory column indicates the primary source of
data (acronyms defined in the text).  ``LAT'' entries jointly fit the indicated parameter family with the GWB signal, e.g.~position and proper motion (``astrometry'') or ``binary'' parameters such as the orbital frequency (FB0) and its time derivatives (FB1, FB2, ...).  When these indicated parameters are fixed instead of fitted, the relative limit on the GWB drops from 1.0 to the value indicated in the fourth column.  The final column indicates pulsars which use fewer than five frequencies to represent stochastic GWB or other red noise signals.  The second, alternative model for PSR~J1959$+$2048 is discussed in the text.}
\vspace{0.3cm}
\begin{tabular}{l | l | l | r | r}
Name & Observatory & LAT Parameters & Change & Harmonics \\
\hline
\hline
PSR~J0030$+$0451 & NRT & -- & -- & --\\
PSR~J0034$-$0534 & NRT & -- & -- & -- \\
PSR~J0101$-$6422 & LAT & binary,astrometry & 0.95 & -- \\
PSR~J0102$+$4839 & NRT & -- & -- & -- \\
PSR~J0312$-$0921 & LAT &  binary,astrometry & 0.86 & 3 \\
PSR~J0340$+$4130 & GBT & -- & -- & -- \\
PSR~J0418$+$6635 & LAT & astrometry & 0.95 & --\\
PSR~J0533$+$6759 & LAT & astrometry & 0.95 & --\\
PSR~J0613$-$0200 & NRT & -- & -- & 2 \\
PSR~J0614$-$3329 & NRT & -- & -- & --\\
PSR~J0740$+$6620 & NRT & -- & -- & --\\
PSR~J1124$-$3653 & LAT &  binary,astrometry,FB1 & 0.92 & --\\
PSR~J1231$-$1411 & NRT & -- & -- & --\\
PSR~J1513$-$2550 & NRT & -- & -- & 2 \\
PSR~J1514$-$4946 & LAT & binary,astrometry & 0.98 & --  \\
PSR~J1536$-$4948 & LAT & binary,astrometry & 0.94 & --\\
PSR~J1543$-$5149 & PKS & -- & -- & 1 \\
PSR~J1614$-$2230 & NRT  & -- & -- & --\\
PSR~J1625$-$0021 & LAT & -- & -- & --\\
PSR~J1630$+$3734 & JBO  & -- & -- & --\\
PSR~J1741$+$1351 & NRT  & -- & -- & 2 \\
PSR~J1810$+$1744 & LAT & binary,astrometry,FB1--8 & 0.73 & -- \\
PSR~J1816$+$4510 & LAT & binary & 0.97 & --\\
PSR~J1858$-$2216 & LAT & binary,astrometry & 0.79 & 1 \\
PSR~J1902$-$5105 & LAT & binary,astrometry & 0.94 & --\\
PSR~J1908$+$2105 & LAT & binary,astrometry & 0.89 & -- \\
PSR~J1939$+$2134 & NRT  & -- & -- & -- \\
PSR~J1959$+$2048 & LAT & position & 0.88 & -- \\
PSR~J1959$+$2048 & LAT & position,binary,FB1--10 & 0.46 & -- \\
PSR~J2017$+$0603 & NRT & -- & -- & --\\
PSR~J2034$+$3632 & LAT & astrometry & 0.87 & 2 \\
PSR~J2043$+$1711 & NRT & -- & -- & --\\
PSR~J2214$+$3000 & NRT & -- & -- & -- \\
PSR~J2241$-$5236 & PKS+LAT & binary & 0.97 & --\\
PSR~J2256$-$1024 & GBT+NRT+LAT & binary & 0.91 & --\\
PSR~J2302$+$4442 & NRT & -- & -- & --\\
\hline
\end{tabular} 
\end{table}

\paragraph*{Ephemerides}
Of the 127 MSPs with 4FGL-DR2 spectral models, only 114 had suitable initial timing solutions, produced using data from the Nan\c{c}ay Radio
Telescope (NRT), Green Bank Observatory (GBT), Arecibo Observatory (AO), the Parkes telescope/Murriyang (PKS), Jodrell Bank Observatory (JBO), and the Giant Metre-wave Radio Telescope (GMRT) (see Table \ref{tab:ephemerides}).   The parameters of these timing solutions were generally well constrained, but in some cases (noted in Table \ref{tab:ephemerides}), the LAT data provided a more precise measurement.  We re-fitted the timing model parameters using a maximum likelihood method, generally finding consistent results with the radio values.  These parameters were allowed to vary in the photon-by-photon GWB analysis, but not in the TOA-based analysis.  $\nu$ and $\dot{\nu}$, the spin frequency and spin-down rate, were allowed to vary in both analyses.

\paragraph*{Solar system ephemeris} In all cases, we used the
DE421 Solar System ephemeris \cite{DE421}, which is commonly used in pulsar timing analyses and facilitates comparison with previous results.
The choice of ephemeris can alter residuals at the $\sim$100\,ns level \cite{Caballero18,Arzoumanian18}, though see \cite{Goncharov21b}, but the LAT data are not yet sensitive to such small effects. 

\paragraph*{Pulse profile templates}
We modeled the pulse profile, $f(\phi)$, using 1--8 Gaussian components wrapped to the interval $0<\phi<1$, which enforces periodicity.  We also fitted these using maximum likelihood, with brighter pulsars having more components.  We held these
templates fixed in the GWB studies described below.


\section*{Photon-by-photon GWB Analysis}

Our photon-by-photon GWB analysis is implemented with maximum likelihood techniques.  Degrees of freedom for spin noise or the GWB are incorporated into the log likelihood,
\begin{equation}
\label{eq:logl_withtn}
\log\mathcal{L}=\sum_i \log \bigg(w_i f(\phi[t_i,\lambda,\beta]) +
(1-w_i)\bigg) -0.5 \sum_{kl} \mathcal{C}_{kl}^{-1}
\beta_k \beta_l -0.5 \det \mathcal{C}.
\end{equation}
The additional timing model parameters $\beta$ are the coefficients of the Fourier transform of a potential noise signal in the data, such that
\begin{equation}
\phi(t,\lambda,\beta)=\phi(t,\lambda)+\sqrt{\frac{2}{t_{\mathrm{obs}}}}\nu^{-1} \sum_k \beta_{2k}\cos\left(2\pi k\frac{t}{t_{\mathrm{obs}}}\right) +\beta_{2k+1}\sin\left(2\pi k\frac{t}{t_{\mathrm{obs}}}\right),
\end{equation}
with $\nu$ the pulsar spin frequency and $t_{\mathrm{obs}}$ the time span of the
data.  The assumption that the noise follows a Gaussian random process is incorporated into the likelihood with the diagonal covariance matrix $\mathcal{C}_{kl}=\delta_{kl}P(f_k,\agwb)/2$ which constrains the amplitudes $\beta$.  $P(f_k)$ is the power spectrum evaluated at the frequency
$f_k=k/t_{\mathrm{obs}}$ \cite{Lentati13}.  For a GWB, theory predicts $P(f) = \frac{\agwb^2}{12\pi^2}\left(\frac{f}{\mathrm{yr}^{-1}}\right)^{-\Gamma=13/3}\,\mathrm{yr}^{-3}$.  In the analysis below we restricted $1\leq k\leq5$ because, as also found by radio PTAs \cite{Arzoumanian20}, we have verified that higher frequencies do not contribute significantly to constraints on the GWB.  For some faint pulsars---typically those with additional free timing model parameters---the data are insufficient to constrain all five harmonics, so we used fewer frequencies as indicated in Table \ref{tab:ephemerides}.  

\begin{figure}
\includegraphics[angle=0,width=0.98\linewidth]{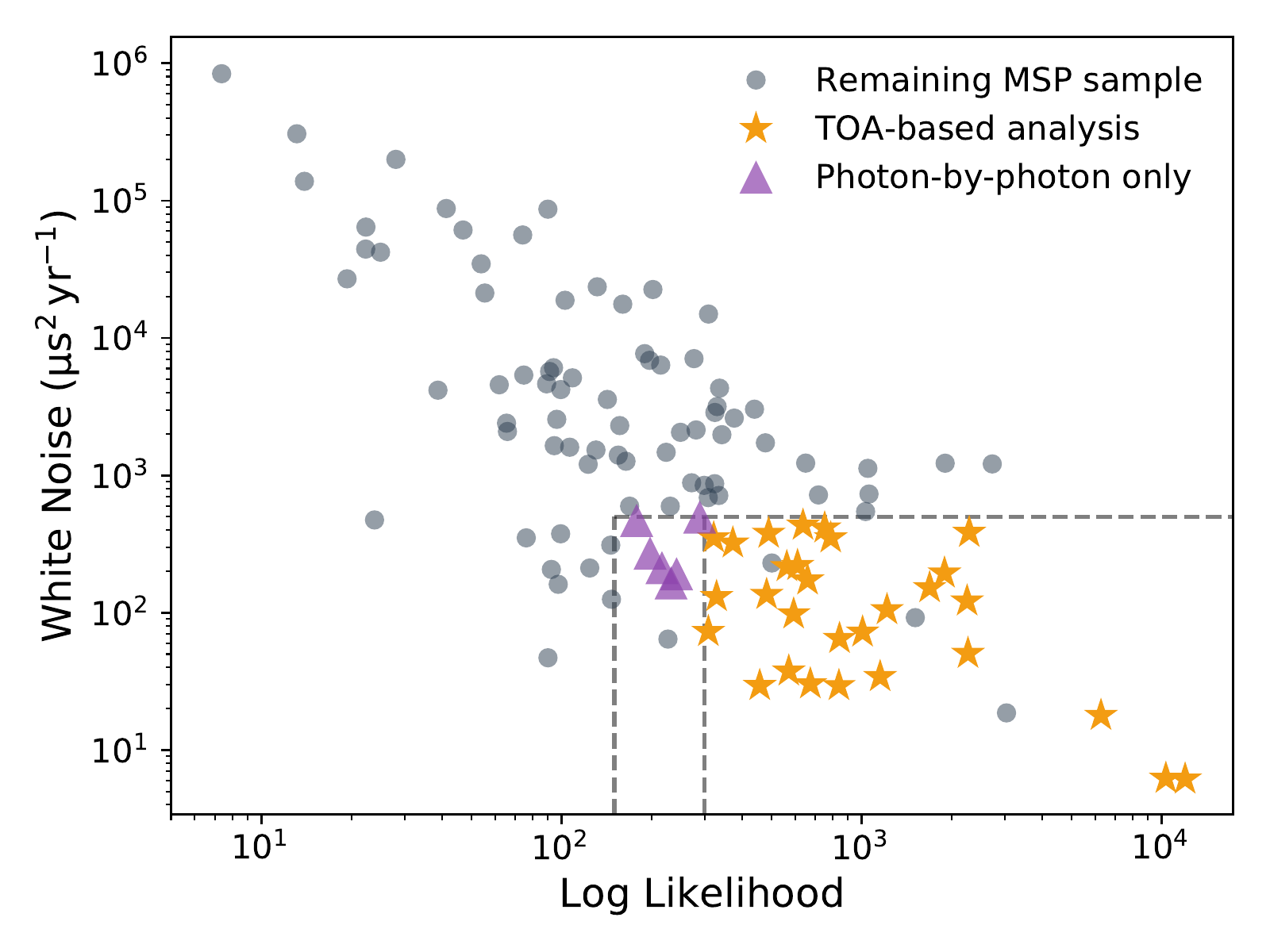}
\caption{\label{fig:wn_vs_logl}\textbf{Timing properties of the 114 MSPs
in the parent sample}. The y axis shows the sensitivity, as estimated by the
limiting amplitude of a white noise process, while the x axis gives
the source significance, or brightness, as estimated by its total log
likelihood, $\log\mathcal{L}$.  Gray dashed lines indicate sample divisions, with the 29 orange stars indicating pulsars used in both analyses and the 6 purple triangles, those MSPs suitable for photon-by-photon analysis only.  Gray circles indicate remaining MSPs.  Four of these---PSR J1311$-$1340, PSR J1555$-$2908, PSR J2215$+$5135, and PSR J2339$-$0533---lie within the selection region but are excluded as discussed in the text.}
\end{figure}

\paragraph*{MSP sample selection}

Only some MSPs are suitable for a GWB analysis.  For each of the 114 pulsars with initial radio timing solutions, we first estimated the white noise level, i.e. the typical amplitude of random fluctuations from the timing model, by averaging the measured amplitudes of 5 harmonics ($\beta_k$ in Equation \ref{eq:logl_withtn}).  In the
absence of other noise, this value is a proxy for sensitivity.  We also calculated the total significance of the pulsed signal via $\log\mathcal{L}$, which is a proxy for the
number of independent TOAs that can be obtained for each pulsar.  These
two values, shown for the full sample in Figure \ref{fig:wn_vs_logl}, are correlated but with
substantial scatter.  At a given intensity, a pulsar with a narrower pulse or faster spin frequency has better timing precision.

Faint MSPs cannot constrain the GWB but would increase computational complexity, so we required a white noise level $<500$\,$\mu$s$^2$\,yr$^{-1}$, corresponding to a typical one-year TOA precision of $\sim$16\,$\mu$s.  We further set $\log\mathcal{L}>150$ for the photon-by-photon
analysis to reduce problems with underconstrained parameters (see below).  In
addition to these requirements, we eliminated four pulsars. PSR~J1311$-$3430 and PSR~J1555$-$2908 have strong spin noise, far in excess of a possible GWB signal, with sufficient amplitude to cause numerical issues in the algorithms described below.  PSR~J2215$+$5135 and PSR~J2339$-$0533 have strong
orbital period variations that require more model components than are supported by \textsc{PINT}. This selection narrowed our sample from 114 pulsars to 35. 

\begin{figure}
\centering
\includegraphics[angle=0,width=0.98\linewidth]{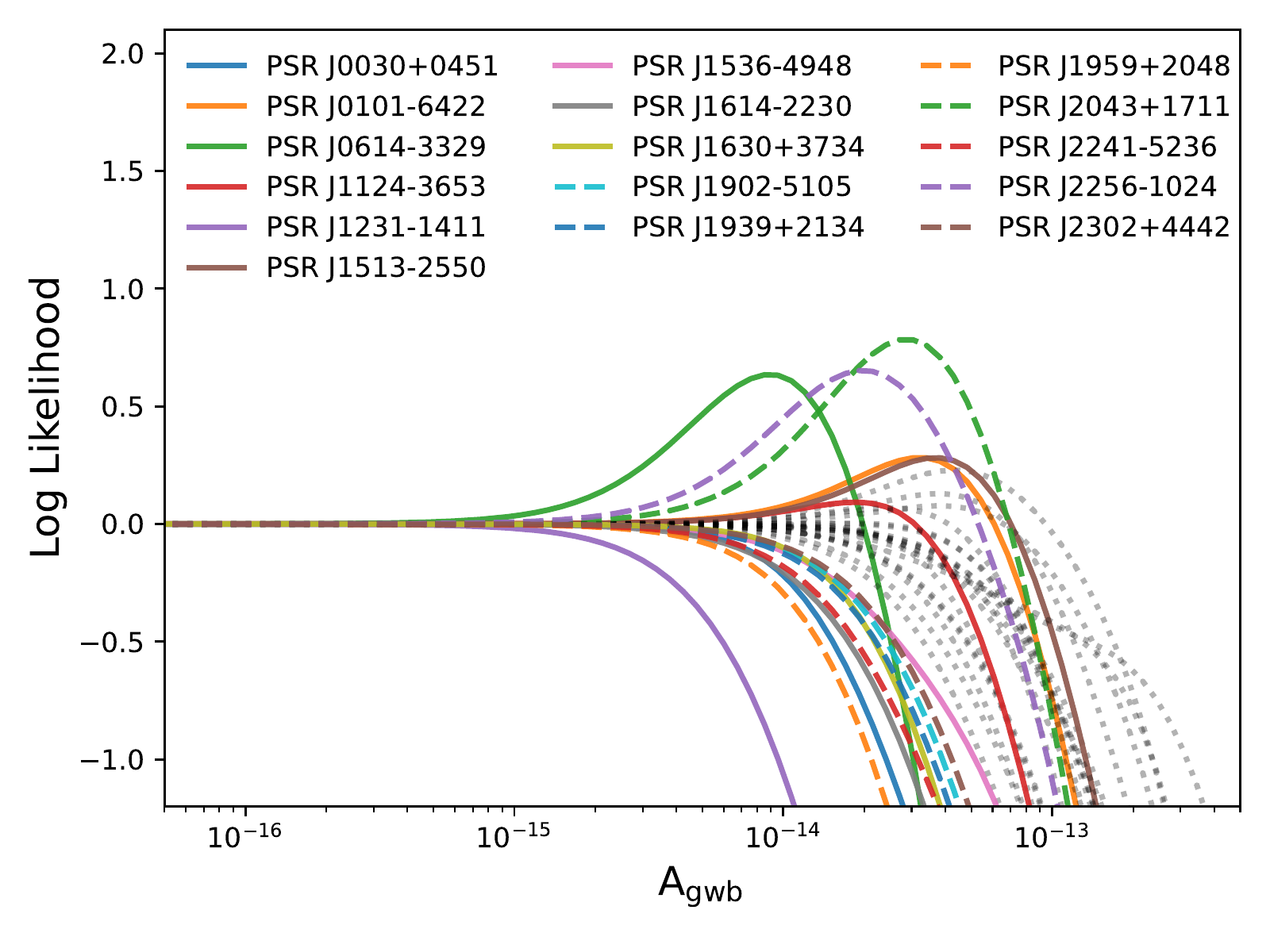}
\caption{\label{fig:unbinned_logl_raw}\textbf{The single-pulsar log likelihoods as a function of $\agwb$ produced by the photon-by-photon method.}  Each likelihood was marginalized analytically over the timing model parameters and Fourier coefficients of the noise process.  The pulsars plotted with colored lines (see legend) have an individual limit $\agwb<1.5\times10^{-13}$ or a peak value at $\agwb>0$ along with a limit $\agwb<3.0\times10^{-13}$.  These are the pulsars with the greatest influence.  The other pulsars are shown in gray dotted lines.}
\end{figure}

\paragraph*{GWB Analysis}
For each pulsar we analyzed a potential GWB signal by evaluating the likelihood following Equation \ref{eq:logl_withtn}, as a function of the timing model parameters $\lambda$ ($\nu$, $\dot{\nu}$, plus any LAT-constrained parameters as indicated in Table \ref{tab:ephemerides}); the Fourier coefficients $\beta$ representing the $\Gamma=13/3$ GWB noise process; and the GWB amplitude $\agwb$.  Remaining timing model parameters were held fixed at radio values; the timing models in our data release \cite{zenodo} provide the exact degrees of freedom for each pulsar.

$\lambda$ and $\beta$ are nuisance parameters. To isolate $\agwb$, we marginalized over them by expanding the log likelihood as a
quadratic form and solving the resulting Gaussian integrals; the likelihood surface for $\beta$ is
generally Gaussian \cite{Kerr19}.  We did this for each pulsar, scanning $\agwb$ over
the range $10^{-20}$--$10^{-10}$. The marginal log likelihoods obtained in this way, as a function of
$\agwb$, are shown in Figure \ref{fig:unbinned_logl_raw}. With a uniform
prior, they are equivalent to the logarithm of the posterior probability distribution.

We obtained single pulsar limits on $\agwb$, listed in Table \ref{tab:singlepulsarlimits}, by integrating the posterior probability density function until accumulating 95\% of it. In no single pulsar is there a strong detection of a GWB-like timing noise process, and in many cases the posterior probability distribution plateaus at $\agwb=0$.  Joint
limits on a common mode are obtained by multiplying the
posteriors and integrating the
resulting distribution.  (This is the ``factorized likelihood'' approach \cite{Arzoumanian20}.)  The models for PSR~J2043$+$1711 and
PSR~J2256$-$1024 are somewhat improved with $\agwb>0$, and when
included in the common mode limit, they substantially increased it.  We
discuss these two pulsars in more detail below.

To gauge the effect of non-GWB degrees of freedom (e.g. proper motion, binary parameters), we also obtained limits with those parameters fixed to their maximum likelihood values.  In
all cases, this improved the resulting limit, typically by 5--10\%
(see Table \ref{tab:ephemerides}), and up to 15--25\% for a few binaries.  In the cases of PSR~J1858$-$2216 (improved by 21\%) and
PSR~J2034$+$3632 (improved by 13\%), the changes can be attributed to improved numerical stability due to the reduced degrees of freedom.  Marked differences also occur for pulsars with
strong variations in the orbital period, PSR~J1810$+$1744 and
PSR~J1959$+$2048, indicating possible degeneracy between these
parameters and the GWB parameters.  Because the time scales associated
with these processes are widely separated, the improved precision could also result from fitting many fewer degrees of freedom, viz. the 8 and 10 orbital frequency derivatives.

\begin{figure}
\centering
\includegraphics[angle=0,width=0.98\linewidth]{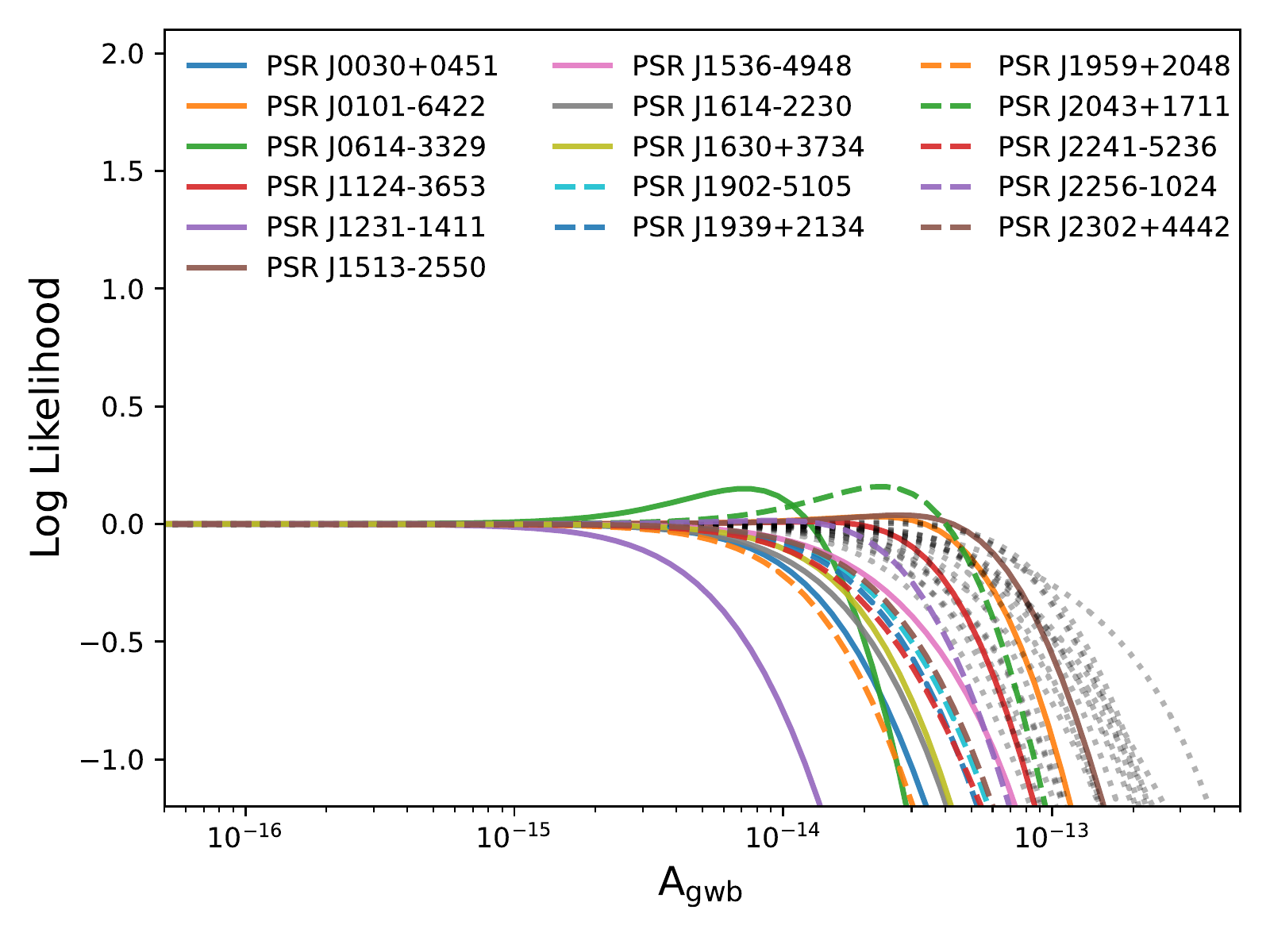}
\caption{\label{fig:unbinned_logl_marg}{\bf The single-pulsar log likelihoods with an additional, numerically marginalized intrinsic spin noise process.}  Other content is as in as Figure \ref{fig:unbinned_logl_raw}.}
\end{figure}

Although we found no strong evidence for spin noise in our sample (see below), we also considered the case of per-pulsar spin noise along with a GWB signal.  For each pulsar, we introduced a general red noise (RN) process with amplitude $A_{\mathrm{RN}}$ and spectral index $\Gamma_{\mathrm{RN}}$, with $A_{\mathrm{RN}}$ defined identically to $A_\mathrm{gwb}$ while $\Gamma_{\mathrm{RN}}$ is allowed to take
values other than $13/3$.  The power spectral density $P_{\mathrm{RN}}(f)$ is also defined identically to $P_{\mathrm{gwb}}(f)$, and we evaluated the likelihood as above but now constraining the Fourier coefficients $\beta$ by the summed power
spectral density $P_{\mathrm{gwb}}(f) + P_{\mathrm{RN}}(f)$.  To marginalize over these two new nuisance parameters, we scanned
over a grid of $A_{\mathrm{RN}}$ and $1<\Gamma_{\mathrm{RN}}<7$ and
tabulated the resulting marginalized log likelihood as a
function of $\agwb$, again using a uniform prior on both $\agwb$ and
$A_{\mathrm{RN}}$.  The log likelihoods in this case are shown in
Figure \ref{fig:unbinned_logl_marg}.  The peaks that appeared in Figure
\ref{fig:unbinned_logl_raw} have reduced or vanished, indicating that
they are likely caused either by statistical fluctuations (i.e. they are
low significance) or by weak timing noise processes with $\Gamma_{\mathrm{RN}}\neq13/3$.  The resulting
constraints on $\agwb$ (Table \ref{tab:singlepulsarlimits}) are generally
poorer, except in cases where the log likelihood with an intrinsic RN
model peaked at $\agwb>0$.  The most conservative limit, reported in the main text, was obtained using this approach.

\section*{TOA-based GWB analysis}
The likelihood based-method described above takes advantage of the time resolution of the LAT data and, by construction, avoids potential systematic errors from reducing the full photon data to TOAs.  However, TOA-based methods are computationally efficient, well-tested, and commonly used by radio PTAs, so we have implemented a parallel TOA-based analysis for comparison. 

\paragraph{TOA estimation} The total log likelihood in Equation \ref{eq:logl_notn} can be interpreted as a measure of
pulsation significance because $f(\phi)=1$ and $\log \mathcal{L}_{data}=0$ in the absence of pulsation.  $\log \mathcal{L}_{data}$ grows as data are accumulated, and once it surpasses a threshold (roughly 20) it is generally possible to reliably measure a phase shift $\delta\phi$ relative to the template.  We converted $\delta\phi$ to a TOA by choosing a reference time near the mid-point of the integration $t_0$ such that $\phi(t_0)\bmod 1=0$ according to a timing model, then iteratively determining $\delta t$ such that $\phi(t_0+\delta_t)\bmod 1=\delta\phi$.  The resulting TOA, $t_0+\delta t$, is in the Universal Time Coordinated (UTC)(GPS) time system because all timestamps are referenced to the on-board GPS clock \cite{Ray11}.  The shape of $\mathcal{L}(\delta\phi)$ becomes more Gaussian as the peak value increases, so a higher $\log\mathcal{L}$ threshold reduces the systematic uncertainty in TOA estimation \cite{Kerr15c} at the expense of a longer integration period.  To minimize non-Gaussian effects, we required $\log\mathcal{L}>300$ to include a pulsar.  For the 29 suitable MSPs, we used the timing models resulting from the photon-by-photon analysis (with any LAT-optimized parameters) and computed TOAs with a cadence of 2, 1.5, and 1 TOAs per year, yielding a total of 25, 19, and 12 TOAs per pulsar, and analyzed (below) each of the data sets before choosing a single representative cadence for the final GWB analysis.

We first applied \textsc{TempoNest} \cite{Lentati14}, which can
constrain a variety of single-pulsar noise models and produce single-pulsar GWB limits
via nested sampling methods.  In particular, we tested for the presence of intrinsic red noise (RN)
and excess white noise (WN).  RN models, like the GWB,
were assumed to be stationary processes with a power-law power spectral
density.  WN models were implemented through the parameters
\texttt{EFAC} and \texttt{EQUAD} (see above).  In general, we expect no excess WN in the gamma-ray data.  However, information is lost when reducing the full photon data to a single TOA and (assumed Gaussian) uncertainty.  This could take the form of increased scatter of TOAs and appear as a WN process.

Table \ref{tab:priorranges} reports the priors on the noise model parameters.  A slightly different range of spectral indices (0--7 vs. 1--7) is adopted between the photon-by-photon and TOA-based analyses, but this has little impact for inference on a steep spectrum process like the GWB.  We used $k=12$ frequency components $f_k$ to model low-frequency noise processes.  Noise process amplitudes follow linear-exponent priors, i.e. the prior probability density $p(A)\propto10^A$, which is equivalent to sampling uniformly from $\log_{10}A$.  All timing model parameters aside from $\nu$ and $\dot{\nu}$ were held fixed.  In all cases, we computed the 
Bayesian evidence for models with and without WN and RN, and the results are presented in 
Table \ref{tab:singlepulsarlimits}. We observed
that the preferred noise model for each pulsar excludes RN and WN
processes for all pulsars except PSR~J1959$+$2048 and PSR~J2241$-$5236.
Both of these pulsars exhibit orbital period variations, and the
excess white noise (detected with very modest Bayes factor $\sim$7), could indicate unmodeled period variations.

\begin{table}
\centering
\caption{\label{tab:priorranges} \textbf{Prior ranges for the noise models used in \textsc{TempoNest} and \textsc{Enterprise}}.  The amplitudes are scaled such that the resulting power spectral density follows Equation \eqgwbamp{}.}
\vspace{0.5cm}
\begin{tabular}{ l | l }
Parameter & Prior ranges \\
\hline
\hline
\texttt{EFAC} & 0.1 to 5      \\
$\log_{10}$ (\texttt{EQUAD}\,s$^{-1}$)  & $-9$ to $-5$      \\
$\log_{10} A_{\mathrm{RN}}$ & $-18$ to $-10$    \\
$\Gamma_{\mathrm{RN}}$ & 0 to 7        \\
$\log_{10} \agwb$ & $-18$ to $-9$    \\
\hline
\end{tabular} 

\end{table}

In general, we found that the results, including the inferred limit on
the GWB, were consistent over the three cadences.  For the faintest
pulsars, compared to longer integrations, we expect the 2\,yr$^{-1}$ cadence to exhibit more systematic errors associated with the possibly-poor Gaussian approximation. This was the case for PSR~J0533$+$6759, PSR~J0740$+$6620, PSR~J1939$+$2134, and PSR~J2034$+$3632, for which models with additional WN were modestly preferred.  With a 1.5\,yr$^{-1}$ cadence, the preferred model for these pulsars required no additional WN.  On the other hand, the highest cadence provides more information about high-frequency noise, so when possible we adopted the 2\,yr$^{-1}$ cadence.  The cadence chosen for each
pulsar is indicated in Table \ref{tab:singlepulsarlimits}.

Using these preferred noise models and cadences, we next used \textsc{Enterprise} \cite{enterprise} to search for GWB signals from each pulsar individually and with a correlation analysis.  The priors on the noise parameters were the same (Table \ref{tab:priorranges}).  Figure \figsinglepulsar{} shows that the \textsc{TempoNest} and
\textsc{Enterprise} limits for individual pulsars are consistent. In the multiple-pulsar analyses, we focused on the simpler common mode process, for which the analysis produced a joint posterior probability distribution for a fixed-amplitude, identical-spectrum noise process with an
independent realization in each pulsar data set.  We also performed a correlation analysis that assumed quadrupolar Hellings-Downs cross-correlation amplitudes \cite{Hellings83}.  These results (see Table \ref{tab:joint_results}) were essentially identical to the common mode analysis. 

\begin{table}
\footnotesize
\centering
\caption{\label{tab:singlepulsarlimits} \textbf{Single pulsar limits on $\agwb$ for 35 pulsars in our sample.} These results use \textsc{TempoNest} (\textsc{TN} in column 4), \textsc{Enterprise} (\textsc{Ent.} in column 5) and the photon-by-photon method (columns 6 and 7). Pulsars with only 
photon-by-photon limits are indicated with an asterisk. Data for PSR~J1959$+$2048 and PSR~J2241$-$5236 favor a model with white noise, while all others favor no additional noise.  Most pulsars can be analyzed with a 2\,yr$^{-1}$ (182\,day) cadence, while six pulsars require longer integrations (1.5\,yr$^{-1}$, 243\,day) to produce reliable TOAs.  The single-pulsar $\agwb$ limits are all 95\% credible levels.  Column 6 (photon-based limits) includes only the GWB in the noise model, while the limits in column 7 also include an intrinsic noise model (RN) for each pulsar, and the parameters of this noise model are marginalized.}
\vspace{0.5cm}
\begin{tabular}{l|r|r|r|r|r|r}
\multicolumn{1}{l|}{Pulsar} & \multicolumn{1}{l|}{Cadence} & Noise model & \multicolumn{1}{l|}{\textsc{TN}} & \multicolumn{1}{l|}{\textsc{Ent.}} & \multicolumn{1}{l|}{Photon} & \multicolumn{1}{l}{Photon+RN}\\
\multicolumn{1}{l|}{} & \multicolumn{1}{r|}{TOA/yr} & \multicolumn{1}{r|}{(favored)} & \multicolumn{1}{r|}{$\times 10^{-14}$} & \multicolumn{1}{r|}{$\times10^{-14}$} & \multicolumn{1}{r|}{$\times10^{-14}$} & \multicolumn{1}{r}{$\times10^{-14}$} \\
\hline
\hline
PSR~J0030$+$0451 & 2 & None & 7.54 & 7.77 & 7.61 & 8.44 \\ 
PSR~J0034$-$0534 & 2 & None & 13.40 & 13.39 & 15.56 & 18.00 \\ 
PSR~J0101$-$6422 & 2 & None & 18.63 & 18.94 & 18.80 &  22.16 \\ 
PSR~J0102$+$4839 & 2 & None & 39.29 & 38.90 & 38.90 & 38.76 \\
PSR~J0312$-$0921* & -- & -- & -- & -- & 21.57 & 27.86 \\ 
PSR~J0340$+$4130 & 2 & None & 26.13 & 26.54 & 48.26 & 58.21 \\ 
PSR~J0418$+$6635* & -- & -- & -- & -- & 32.59 & 36.41 \\ 
PSR~J0533$+$6759 & 1.5 & None & 21.66 & 22.14 & 21.83 & 26.23\\ 
PSR~J0613$-$0200 & 2 & None & 22.57 & 21.59 & 21.66 & 25.80 \\ 
PSR~J0614$-$3329 & 2 & None & 4.15 & 4.20 & 3.94 & 4.36 \\ 
PSR~J0740$+$6620 & 1.5 & None & 15.76 & 16.62 & 17.24 & 20.55 \\ 
PSR~J1124$-$3653 & 2 & None & 15.84 & 15.58 & 14.34 & 16.77 \\ 
PSR~J1231$-$1411 & 2 & None & 2.19 & 2.30 & 2.71 & 3.54 \\ 
PSR~J1513$-$2550* & -- & -- & -- & -- & 25.57 & 43.22 \\
PSR~J1514$-$4946 & 2 & None & 38.89 & 38.54 & 41.85 & 34.90 \\
PSR~J1536$-$4948 & 2 & None & 12.30 & 11.69 & 14.36& 15.38 \\
PSR~J1543$-$5149* & -- & -- & -- & -- & 98.02 & 1356.47 \\
PSR~J1614$-$2230 & 2 & None & 9.08 & 9.23 & 8.14 & 9.76 \\
PSR~J1625$-$0021 & 1.5 & None & 30.70 & 30.74 & 28.01 & 31.07\\
PSR~J1630$+$3734 & 2 & None & 7.28 & 7.27 & 7.91 & 9.08\\
PSR~J1741$+$1351* & -- & -- & -- & -- & 84.24 & 120.30\\
PSR~J1810$+$1744 & 2 & None & 14.31 & 14.95 & 18.54 & 21.35 \\
PSR~J1816$+$4510 & 2 & None & 34.91 & 35.61 & 39.09 & 41.31\\
PSR~J1858$-$2216 & 2 & None & 31.51 & 30.46 & 110.83 & 1417.63 \\
PSR~J1902$-$5105 & 2 & None & 11.50 & 11.38 & 11.87 & 15.06\\
PSR~J1908$+$2105* & -- & -- & -- & -- & 41.80 & 47.97 \\
PSR~J1939$+$2134 & 1.5 & None & 12.86 & 13.04 & 10.24 & 12.99 \\
PSR~J1959$+$2048 & 2 &   WN & 8.25 & 8.12 & 6.15 & 7.84 \\
PSR~J2017$+$0603 & 2 & None & 16.63 & 16.60 & 17.59 & 20.25 \\
PSR~J2034$+$3632 & 1.5 & None & 21.62 & 22.21 & 38.82 & 74.75\\
PSR~J2043$+$1711 & 2 & None & 13.38 & 13.80 & 13.96 & 15.07\\
PSR~J2214$+$3000 & 2 & None & 29.67 & 30.44 & 28.74 & 38.60 \\
PSR~J2241$-$5236 & 2 & WN & 13.80 & 13.66 & 14.05 & 16.39 \\
PSR~J2256$-$1024 & 1.5 & None & 12.98 & 13.23 & 13.84 & 13.31 \\
PSR~J2302$+$4442 & 2 & None & 11.74 & 11.67 & 12.26 & 14.82 \\
\hline
\end{tabular}%
\end{table}

\section*{Results}

\paragraph*{Comparison of methods}

The two codes \textsc{Enterprise} and
\textsc{TempoNest} provided single pulsar limits which were consistent with each other, so we compared TOA-based and photon-by-photon approaches.  Aside from computational aspects---the TOA-based methods are sampled, while the photon-by-photon method is analytic---there are two primary differences: the
photon-based approach avoids the assumption of Gaussianity on
TOA uncertainties, and it retains sensitivity to all timescales.  We expect the photon-based approach to generally be more precise due to these advantages.  The agreement between these two methods provides us with confidence in our photon-by-photon approach. The few exceptions (discussed below) stem from the fundamental differences in the methods, and they have little impact on the final GWB limits because none of these pulsars contribute strongly to the sensitivity of the timing array.

PSR~J0340$+$4130 was the largest outlier, with a photon-by-photon 
limit nearly twice that of the TOA-based methods.  The degrees of freedom in the two methods are identical, but the photon-by-photon log likelihood peaks at $\agwb>0$, suggesting a real fluctuation or weak timing noise.  We speculate non-Gaussianity may coincidentally distort the TOAs for this pulsar in such a way as diminish this possible noise and thus reduce the GWB limit.

For PSR~J1536$-$4948, the photon-based method also delivered a higher limit, probably because the photon-by-photon model includes free binary and astrometry parameters. The compact binary PSR~J1810$+$1744 has extensive orbital period variations, and these degrees of freedom are likewise inaccessible to
the TOA-based method.  When we fixed these binary parameters, the photon-by-photon dropped by $\sim$30\% and agreed with the TOA-based results.

PSR~J1858$-$2215 and PSR~J2034$+$3632 are both faint pulsars, so our models used a
restricted number of harmonics for numerical stability (see above).  This degraded the sensitivity of the photon-based analysis.

\paragraph*{Combined limits and scaling}
We combined the single-pulsar limits discussed above to obtain an overall limit on the GWB amplitude.  To assess the dependence of the total limit on particular pulsars, we obtained results for both the full pulsar timing array and for some subsets.
The best nine pulsars in ascending order of single pulsar upper limits are PSR~J1231$-$1411, PSR~J0614$-$3329, PSR~J1959+2048, PSR~J0030+0451, PSR~J1630+3734, PSR~J1614$-$2230, PSR~J1939+2134, PSR~J1902$-$5105 and PSR~J2302+4442. These pulsars form the subsets listed in Table \ref{tab:joint_results}.
For both methods, the limits steadily improved with additional pulsars.  In the case of the TOA-based approach, the limit also improved with the inclusion of the spatial correlation information predicted for a GWB \cite{Hellings83}.  The two full-array limits were nearly identical, while the tightest constraint was $\agwb<1.0\times10^{-14}$.

\begin{table}
\centering
\caption{\label{tab:joint_results} \textbf{95\% credible upper limits
on $\agwb/10^{-14}$ from the combined samples.}  The pulsars corresponding to each subset are specified in the text and are ranked by their single-pulsar GWB upper limits.  The ``Full'' rows indicate the total sample for the two methods, 29 pulsars common to TOA-based and photon-by-photon, and 35 to photon-by-photon only.  The final row includes the Hellings-Downs (HD) spatial correlations predicted for quadrupolar gravitational radiation \cite{Hellings83}.}
\vspace{0.5cm}
\begin{tabular}{l | r | r | r | r | r }
\multicolumn{1}{l|}{Subset} & \multicolumn{1}{l|}{\textsc{Enterprise}} & \multicolumn{1}{l|}{\textsc{Enterprise}} & \multicolumn{1}{l|}{Photon} & \multicolumn{1}{l|}{Photon} \\
\multicolumn{1}{l|}{} & \multicolumn{1}{l|}{} & \multicolumn{1}{l|}{with RN} & \multicolumn{1}{l|}{} & \multicolumn{1}{l|}{with RN} \\
\hline
\hline
Best 2        & 1.89 & 1.94 & 1.84 & 2.02 \\
Best 3        & 1.71 & 1.74 & 1.50 & 1.66 \\
Best 9        & 1.15 & 1.19 & 1.06 & 1.16 \\
Full 29       & 1.12 & 1.04 & 1.14 & 1.06 \\
Full 35       &  --  &  --  & 1.14 & 1.05 \\
Full 29 w/HD     & 1.06 & --   & -- & --\\

\hline 
\end{tabular} 

\end{table}

As discussed above, the limit in the photon-by-photon case can be degraded by including pulsars whose $\agwb$ posteriors peak at $\agwb>0$, which can happen even in the absence of a GWB signal due to statistical fluctuations.  For completeness, however, we considered limits computed when removing two MSPs with the strongest signals, PSR~J2043$+$1711 and PSR~J2256$-$1024 (which exhibits modest orbital period variations).  This is justifiable if the log likelihood peak is due to some deficiency in the timing solution.  When removing these two pulsars, we obtained limits of $9.8\times10^{-15}$ and $10.3\times10^{-15}$.

We additionally considered any systematic effects from our choice of the DE421 Solar System ephemeris, by performing the same limit calculation using a perturbative Bayesian modeling software, \textsc{BayesEphem} \cite{Vallisneri2021}. This allowed us to model uncertainties in the Solar System planetary masses and orbital parameters while simultaneously constraining other pulsar noise parameters and the GWB.  The \textsc{Enterprise} results, obtained with and without \textsc{BayesEphem}, are reported in Table \ref{tab:joint_bayes}.  The limits with \textsc{BayesEphem} were slightly tighter for the few-pulsar subsets, which follows from the increased degeneracy between the GWB quadrupolar signature and dipolar-like effects from Solar System ephemeris errors \cite{Tiburzi16}.  The full array results, on the other hand, contained enough information to separate the two signals.  The agreement between our limits fixed at DE421 and those obtained with \textsc{BayesEphem} indicates its use does not bias our results.

\begin{table}
\centering
\caption{\label{tab:joint_bayes}\textbf{95\% credible upper limits
on $\agwb/10^{-14}$ from the combined samples using \textsc{BayesEphem}}. Here, we compare results obtained with \textsc{Enterprise}, in both cases using spatial correlations \cite{Hellings83} but with and without \textsc{BayesEphem}.}
\vspace{0.5cm}
\begin{tabular}{l | r | r | r }
\multicolumn{1}{l|}{Subset} & \multicolumn{1}{l|}{\textsc{Enterprise}} & \multicolumn{1}{l|}{\textsc{Enterprise}} \\
\multicolumn{1}{l|}{} & \multicolumn{1}{l|}{} & \multicolumn{1}{l|}{with \textsc{BayesEphem}} \\
\hline
\hline
Best 2        & 1.90 & 1.81 \\
Best 3        & 1.70 & 1.64 \\
Best 9        & 1.17 & 1.16 \\
Full 29       & 1.06 & 1.08 \\

\hline 
\end{tabular} 

\end{table}

\begin{table}
\centering
\caption{\label{tab:pub_limits}\textbf{Published radio PTA results on the GWB and GWB-like signals}.  Dates given are fractional year of publication.  Upper limits (u.l.) and central values are given in the ``$A_{\mathrm{gwb}}$'' column while the ``Range'' column gives central ranges of GWB-like candidate signals.  These values are plotted in Figure 1.}
\vspace{0.5cm}
\begin{tabular}{l | l | r | r | r | l  }
\multicolumn{1}{l|}{Label} & \multicolumn{1}{l|}{Year} & \multicolumn{1}{l|}{Reference} & \multicolumn{1}{l|}{$\agwb$} & \multicolumn{1}{l|}{Range} & \multicolumn{1}{l}{Note}\\
\multicolumn{1}{l|}{} & \multicolumn{1}{l|}{}  & \multicolumn{1}{l|}{} & \multicolumn{1}{l|}{$\times10^{-15}$} & \multicolumn{1}{l|}{$\times10^{-15}$} &  \\
\hline
\hline
PPTA 2006 & 2006.96 & \cite{Jenet06}         & 11  & -- & 95\% u.l. \\
PPTA 2013 & 2013.79 & \cite{Shannon13} & 2.4 & -- & 95\% u.l. \\
PPTA 2015 & 2015.71 & \cite{Shannon15}       & 1.0 & -- &  95\% u.l. \\
PPTA 2021 & 2021.54 & \cite{Goncharov21b}  & 2.2 & 1.9--2.6 & 68\% range \\
EPTA 2011 & 2011.54 & \cite{vanHaasteren11}  & 6.0 & -- & 95\% u.l. \\
EPTA 2015 & 2015.88 & \cite{Lentati15}       & 3.0 & -- & 95\% u.l. \\
EPTA DR2  & 2021.96 & \cite{Chen21} &-- & 2.3--3.7 & 5--95\% credible region \\
NANOGrav 5-yr & 2013.04 & \cite{Demorest13} & 7.0 & -- & 95\% u.l. \\
NANOGrav 11-yr & 2018.38 &\cite{Arzoumanian18} & 1.45 & -- & 95\% u.l.\\
NANOGrav 12.5-yr & 2020.96 & \cite{Arzoumanian20} & 1.92 & 1.4--2.7 & 5--95\% credible region \\
IPTA DR1 & 2016.38 & \cite{Verbiest16} & 1.7 & -- & 95\% u.l.\\ 
IPTA DR2 & 2022.21 & \cite{Antoniadis22} & -- & 2.0--3.6 & 5--95\% credible region \\
\hline 
\end{tabular} 

\end{table}

In summary, we obtained an upper limit of $\agwb<10^{-14}$, with two independent methods yielding both expected scalings and consistent values.  The limit was unaffected by the Solar System ephemeris (DE421) we chose, and to particular realizations of noise modeling, e.g.~the inclusion of per-pulsar spin noise.

\paragraph{Other GWB models}
 Using the photon-by-photon method, we calculated 95\% upper limits on more general power-law GWB models, shown in Figure \figagwbvsindx{}.  For each value of $\alpha$, we computed a photon-based limit as described above, without including any per-pulsar spin noise.  The resulting scaling is approximately exponential, and the scaling largely reflects the fact that the limit is dominated by frequencies around $2/t_{\mathrm{obs}}$.  Limits are slightly less constraining for models with more high-frequency power due to the limited number of Fourier coefficients used in the analysis.

\paragraph{Spin noise}
There was little evidence for intrinsic spin noise in any of the pulsars in our sample (Table \ref{tab:singlepulsarlimits} and Figure \ref{fig:unbinned_logl_raw}).  PTAs have measured such intrinsic timing noise for some of them, but the amplitudes are generally below the single-pulsar sensitivity. PSR~J1939$+$2134 has intrinsic timing noise evident over the decades-long data sets \cite{Verbiest16}.  We detect no spin noise and obtained limits somewhat below the values reported by radio PTAs (see below).

Although we saw no strong evidence for spin noise in any pulsar, PSR~J2043$+$1711 and PSR~J2256$-$1024 had posterior probability distributions which peaked at $\agwb>0$, but at sufficiently low values of $\agwb$ that the combined photon-by-photon limit increases.  The posterior peaks decreased when we included a spin noise model and marginalized over its parameters, indicating the possible noise is likely more consistent with a different spectral index.  In general, the single-pulsar $\agwb$ limits increased when including these extra degrees of freedom (Table \ref{tab:singlepulsarlimits}), but decreased for pulsars with possible intrinsic RN.  The total limit on $\agwb$ is slightly lower when including spin noise, but both values were within 10\% of each other.

\paragraph*{Comparison to radio measurements}

\begin{table}
\centering
\caption{\label{tab:radio_comp}\textbf{Comparison between measured radio PTA spin noise amplitudes and 95\% Fermi PTA upper limits.} The spin noise power spectra are given in s$^2$\,yr$^{-1}$ evaluated at $f=1/\mathrm{yr}$.  Fermi limits are computed similarly to GWB amplitude limits except using the value of $\Gamma$ measured in each row rather than $\Gamma=13/3$.  The last columns give the ratio of the Fermi upper limit to the radio PTA power.}
\vspace{0.5cm}
\begin{tabular}{l | r | r | r | r | r }
\multicolumn{1}{l|}{Pulsar} & \multicolumn{1}{l|}{PTA} &  \multicolumn{1}{l|}{$\Gamma$} & \multicolumn{1}{l|}{P(f) PTA} & \multicolumn{1}{l|}{P(f) Fermi} & \multicolumn{1}{l}{Ratio}\\
\hline
\hline
PSR J0030$+$0451 & NANOGrav &  6.3 & $9.0\times10^{-18}$ & $8.3\times10^{-18}$ & 0.9 \\
PSR J0613$-$0200 & NANOGrav &  2.1 & $1.5\times10^{-14}$ & $2.0\times10^{-13}$ & 13 \\
PSR J0613$-$0200 & PPTA &  4.2 & $2.5\times10^{-16}$ & $5.4\times10^{-15}$ & 21 \\
PSR J1939$+$2134 & NANOGrav &  3.3 & $9.8\times10^{-15}$ & $6.6\times10^{-15}$ & 0.7 \\
PSR J1939$+$2134 & PPTA &  5.4 & $1.8\times10^{-16}$ & $1.1\times10^{-16}$ & 0.6 \\

\hline 
\end{tabular} 

\end{table}

Radio PTAs have published spin noise models for many MSPs, particularly two recent sets based on the NANOGrav 12.5-yr data set \cite{Alam21} and the PPTA data release 2 \cite{Kerr20,Goncharov21a} containing measurements for 14 and 10 MSPs, respectively.   Both contain treatments for the time-varying DM (NANOGrav use \texttt{DMX} and the PPTA use a stationary process model), so these should be unbiased measurements of the spin noise.  Of these pulsars, only 3 are in common with our $\gamma$-ray MSP sample, PSR~J0030$+$0451, PSR~J0613$-$0200, and PSR~J1939$+$2134.  For these pulsars, the radio estimates for spin noise do differ, with the PPTA analysis finding  steeper spectra for the two common pulsars in our sample.

We computed 95\% upper limits on a spin-noise process for each pulsar using the radio-PTA measured spectral index $\Gamma$, with results listed in Table \ref{tab:radio_comp}.  Aside from the changed value of $\Gamma$, the procedure is identical to the photon-by-photon method we used to compute single-pulsar GWB limits.  The Fermi upper limits for PSR~J0030$+$0451 and PSR~J1939$+$2134 were lower than the measured radio values and statistically incompatible.  This could be evidence for uncorrected IISM/solar wind effects leaking into the spin noise estimation in the radio data.  As discussed above, radio emission from PSR~J0030$+$0451 is particularly affected by the solar wind because its position passes near to the Sun each year.

Although the number of MSPs here is too small to draw firm conclusions, there is some evidence for contamination of spin noise models (and, ultimately, GWB-like signals) by uncorrected IISM effects.  The most sensitive gamma-ray MSPs, PSR~J1231$-$1411, PSR~J0614$-$3329, and others are not among the pulsars routinely monitored by radio PTAs, so we cannot compare further.

\end{document}